\documentclass[reprint,superscriptaddress, amsmath,amssymb, aps]{revtex4-2}

\usepackage{xcolor}
\usepackage{graphicx}
\usepackage{float}
\usepackage{dcolumn}
\usepackage{bm}
\usepackage{cancel}

\usepackage{url}

\begin{document}

\preprint{APS/123-QED}

\title{Endemic infectious states below the epidemic threshold and beyond herd immunity
}

\author{Javier Aguilar}
\author{Beatriz Arregui García}
\author{Ra\'ul Toral}
\author{Sandro Meloni}\email{sandro@ifisc.uib-csic.es}
\author{Jos\'e J. Ramasco}

\affiliation{Instituto de F\'{\i}sica Interdisciplinar y Sistemas Complejos IFISC (CSIC-UIB), Campus UIB, 07122 Palma de Mallorca, Spain. }

\date{\today}
\begin{abstract}
In the recent COVID-19 pandemic we assisted at a sequence of epidemic waves intertwined by anomalous fade-outs with periods of low but persistent epidemic prevalence. These long-living epidemic states complicate epidemic control and challenge current modeling approaches as classical epidemic models fail to explain their emergence. 
Inspired by this phenomenon, we propose a simple mechanism able to reproduce several features observed in real data. Specifically, here we introduce a modification of the Susceptible-Infected-Recovered (SIR) model in a meta-population framework where a small inflow of infected individuals accounts for undetected internal or imported cases. Focusing on a regime where this external seeding is so small that cannot be detected from the analysis of epidemic curves, we find that outbreaks of finite duration percolate in time resulting in overall low but long-living epidemic states below and above the epidemic threshold. Using a two-state description of the local dynamics, we can extract analytical predictions for the phase space. The comparison with epidemic data demonstrates that our model is able to reproduce some critical signatures observed in COVID-19 spreading in England. 
Finally, our findings defy our understanding of the concept of epidemic threshold and its relationship with outbreaks survival for disease control.  
\end{abstract}

\maketitle

\section{Introduction}
\label{Intro}
The proliferation of infectious diseases is inherent to the social condition of human beings, and it has strongly marked cultural evolution in the last millennia~\cite{campbell2005tuberculosis,bigon2012history,banai2020pandemic}. Certainly, our current bio-chemical knowledge is mature enough to offer pharmaceutical solutions for many diseases. However, the structure of our societies and our way of living (e.g. rapid communications, highly connected world, dense urban areas, pollution, climate change, etc.) contribute to the appearance and quick diffusion of new health threats ~\cite{chinazzi2020effect}. Indeed, the experience of the COVID-19 pandemic highlighted the importance of understanding other aspects of infectious diseases spreading such as: the effect of non-pharmaceutical interventions~\cite{haug2020ranking,perra2021non,oh2021mobility}, long-range travel restrictions~\cite{chinazzi2020effect}, the predictability of epidemic models~\cite{castro2020turning}, the impact of city structure~\cite{aguilar2020impact,arenas2020modeling} and their effect on  public opinion~\cite{gallotti2020assessing,briand2021infodemics,PhysRevResearch.4.013158}, etc.


In this context, we focus here on the anomalous behavior of COVID-19 fade-outs. It was observed that epidemic curves after the first epidemic wave are characterized by oscillations, plateaus, linear growth of the total number of cases and high susceptibility to secondary waves~\cite{weitz2020awareness,tkachenko2021time,neri2021role,thurner2020network,wu2020covid,maier2020effective}. These phenomena result in overall long-living, yet marginal, endemic states that differ from the classical exponential decay that one would naively expect. 
This makes difficult to assess when control measures could be relaxed since their lifting may lead to new major outbreaks. 
Furthermore, long survival times can easily result in ``epidemic broths'' where new variants can emerge. Therefore, a proper evaluation of the plausible causes of these long-living states is fundamental to design interventions aimed at controlling disease spreading. 


Since classical models fail in predicting persistent small fluctuations close to absorbing states, this situation strikes our understanding of disease spreading and poses a fundamental problem for disease modeling: which is the minimal epidemic model able to generate the observed phenomenology?
This topic is widely discussed in the literature. While some works assert that this non-trivial temporal behavior is the result of the intrinsic heterogeneity in the infection parameters or in the structure of the contact networks~\cite{berestycki2021effects,tkachenko2021time,neri2021role,thurner2020network}, another line of research suggests that individual response could be at its basis, and, in turn, it may depend on the state of the disease~\cite{wu2020covid,weitz2020awareness,maier2020effective}. Moreover, the physics community also noted how this phenomenology reminds of the features observed at a fined-tuned critical point. For example, \cite{Radicchi2020} shows that the linear growth of the total number of cases and incidence plateaus can be induced by the initial conditions of the Susceptible-Infected-Recoved (SIR) model at the critical point while other authors are looking for self-organization mechanisms in order to avoid the dependence on fine-tuned parameters~\cite{ariel2021self,manrubia2021individual}.
 
In this work, we show that the presence of a small number of undetected cases, either coming from importation or due to local missed detection, explains the emergence of these anomalous fade-outs and long-living states. 
To show the effect of this mechanism, we study a meta-population epidemic model with a small external seeding. Our approach  does not depend, thus, on ad-hoc model modifications, complex behavioral modeling, fine-tuning, or self-organization mechanisms. By means of a coarse-grain of the epidemic dynamics, we are able to extract analytical information about the duration of outbreaks after the first wave. This is a novel analytical procedure to extract global information out of local properties in the context of meta-population models. Our results confirm that endemic states could be sustained by the minimal import of infected individuals below and above the epidemic threshold. This shows that driving epidemics just below the basic reproductive number may not result in a total epidemic fade-out. Moreover, it also makes manifest that the whole concept of herd immunity should be revisited. Finally, we also confront our theoretical derivations with empirical epidemic data of COVID-19 in England, finding that our mechanism can explain the anomalous persistence of the disease and the observed signatures of criticality after the end of the first wave.

 

\section{External forcing}\label{sec:external_forcing}   

By external seeding (forcing), we mean a process that introduces newly infected individuals to a population. We mainly consider the regime in which these arrivals follow a slow rate, in such a way that the external forcing can be seen as a small perturbation on the system dynamics. Of course, in the very early moments and depending on the epidemic parameters, seeding has the potential to trigger an outbreak. However, rather than in the first wave we are interested in the effects of  forcing in later stages of the epidemic spreading. 

Such external seeding can be a description of a myriad of processes. An example could be the effect in a certain region or country of trips of undetected infected individuals from the rest of the world, which would act as a reservoir. This can be combined with other mechanisms such as failures in the isolation of infected people, the effect of asymptomatic individuals traveling, false negative tests, non-perfect mobility restrictions, etc. In general, there is no epidemic control strategy that is infallible. The intention is to model all these undetected cases as an effective (very low) inflow of epidemic seeds. 

The effect of external seeding on a single-population SIR model has been studied in detail (see e.g.~\cite{singh2014outbreak,stollenwerk2021interplay}). Also, the relevance of meta-population models to study scenarios with realistic topologies such as cities, countries, or global airport connections has been extensively explored (e.g. \cite{grais2003assessing,sattenspiel1995structured,driessche2008spatial,rvachev1985mathematical,colizza2007reaction,balcan2009multiscale,balcan2009seasonal,ajelli2010comparing,tizzoni2012real,aguilar2020impact}). For instance, a recent work has focused on the effects of multiseeding on a meta-population framework~\cite{mazzoli2021}. However, the seeding, in that case, was not external, since it originated in other sub-populations, and it was not particularly small or constant over time. The question that remains open and that we address in this paper is the impact of this small external seeding from a reservoir in a meta-population, especially, in the period between epidemic waves. 

\section{Single population SIR model with external seeding} \label{sec:one_pop}

In a single well-mixed population, the SIR model with external seeding is defined by the following rules: Infected individuals become Recovered at rate $\mu$, Susceptible individuals are infected after contact with an infected agent at rate $\beta$. Lastly, a random individual can be substituted by an external infected agent with rate $h$. We use substitution, instead of direct introduction, to conserve the total population $N$ constant ($N=S+I+R$, where $S$, $I$ and $R$ are the number of Susceptible, Infected and Recovered individuals, respectively). In our approach, we focus on continuous time stochastic models that generate different epidemic curves [$I(t),S(t),R(t)$] in every realization.
These rules are encoded as the transition rates of an stochastic Markov jumping process~\cite{jacobs2010stochastic}:
\begin{align}\label{eq:macroscopic_rates_ATA_SIR_plus_field}
    & \lim_{dt\to 0}\frac{P(I+1,S-1,R,t+dt|I,S,R,t)}{dt}=\beta \frac{I}{N} S, \nonumber \\
    & \lim_{dt\to 0}\frac{P(I-1,S,R+1,t+dt|I,S,R,t)}{dt}=\mu I, \nonumber
    \\
    & \lim_{dt\to 0}\frac{P(I+1,S-1,R,t+dt|I,S,R,t)}{dt}=\frac{h}{N}S , \nonumber \\
    & \lim_{dt\to 0}\frac{P(I+1,S,R-1,t+dt|I,S,R,t)}{dt}=\frac{h}{N} R.
\end{align}
Here the seeding rate appears in the form $h/N$ to account for the substitution of a small number of individuals per unit of time. If instead, $h$ multiplied $S$ and $R$ directly, it would represent the substitution of a fraction of the total population. This further reinforces our message that we are considering small external seeding.

The dynamics of the system in the limit of large population ($N\rightarrow \infty$) can be approximated by the set of deterministic equations:
\begin{align}
\label{eq:SIRh_MF}
 \frac{dS}{dt}&=-\beta \, I\, \frac{S}{N}- \frac{h}{N}S , \nonumber \\
 \frac{dI}{dt}&= \beta \, I \, \frac{S}{N} - \mu\, I+\frac{h}{N}(S+R) , \\
 \frac{dR}{dt}&= \mu \, I- \frac{h}{N}R , \nonumber
\end{align}
We proceed to summarize some important information derived from Eqs.~\eqref{eq:SIRh_MF} that will help us to better understand the stochastic model that will be considered afterward.

\subsection{Absence of external forcing ($h=0$)}

When $h=0$, the dynamics reduces to the classical SIR model and any state with $I=0$ is an absorbing fixed point. The behavior towards the absorbing state is controlled by the basic reproductive number $\mathcal{R}_0=\beta/\mu$. If $\mathcal{R}_0>1$, the system is in the super-critical phase, characterized first by an exponential growth of the infected individuals and a subsequent exponential decrease, once the number of susceptible individuals is so low that it cannot fuel the epidemic spreading. This passive phenomenon based on starving out the epidemic spread thanks to the development of an immune community is usually called \emph{herd immunity}. Contrary, if $\mathcal{R}_0<1$, the mean-field equations predict a monotonic exponential decay of the number of infected individuals. The case $\mathcal{R}_0=1$ is then the critical point, separating the super-critical and sub-critical phases.

These two phases also differ in their stationary states. Whereas  the disease reaches a macroscopic fraction of the population ($\lim_{t\rightarrow \infty} R(t)\sim \mathcal{O}(N)$) when $\mathcal{R}_0>1$, the sub-critical ($\mathcal{R}_0<1$) fraction of affected individuals will be small ($\lim_{t\rightarrow \infty} R(t)\sim \mathcal{O}(1)$). This fact allows us to define the \emph{attack rate}~\cite{colizza2008epidemic},
\begin{equation}
\alpha:=\lim_{t\rightarrow \infty}\frac{R(t)}{N},
\end{equation}
as a control parameter. The attack rate in the SIR model can be computed exactly, and its analytical expression will be useful in the derivations of section~\ref{sec:independent_pop}:
\begin{equation}\label{eq:attack_rate}
    \alpha=1+\mathcal{R}_0^{-1}\, \mathcal{W}\left(-s_0\, \mathcal{R}_0\, e^{-\mathcal{R}_0}\right).
\end{equation}
Where $s_0=\frac{S(0)}{N}$ is the initial fraction of susceptible individuals and $\mathcal{W}(\cdot)$ is the Lambert function. See appendix~\ref{ap:attack_rate_SIR} for proof of the above equation.
The critical point is characterized both by a null attack rate ($\alpha = 0$), together with a linear growth of the recovered individuals $(R(t)\propto t)$~\cite{Radicchi2020}.

Not only the magnitude of the outbreak, but also its duration is greatly determined by the basic reproductive number $\mathcal{R}_0$. In the case of the stochastic finite-population SIR model, every realization of the process will have a different value of the attack rate and a different duration. However, Eq.~\eqref{eq:attack_rate} will still be representative of its average behavior. The stochastic nature is of special relevance close to the critical point, where there is a dominance of fluctuations, with a strongly varying number of new cases per unit of time~\cite{marro2005nonequilibrium,barrat2008dynamical,henkel2008non}.

\begin{figure*}
 \includegraphics[scale=0.3]{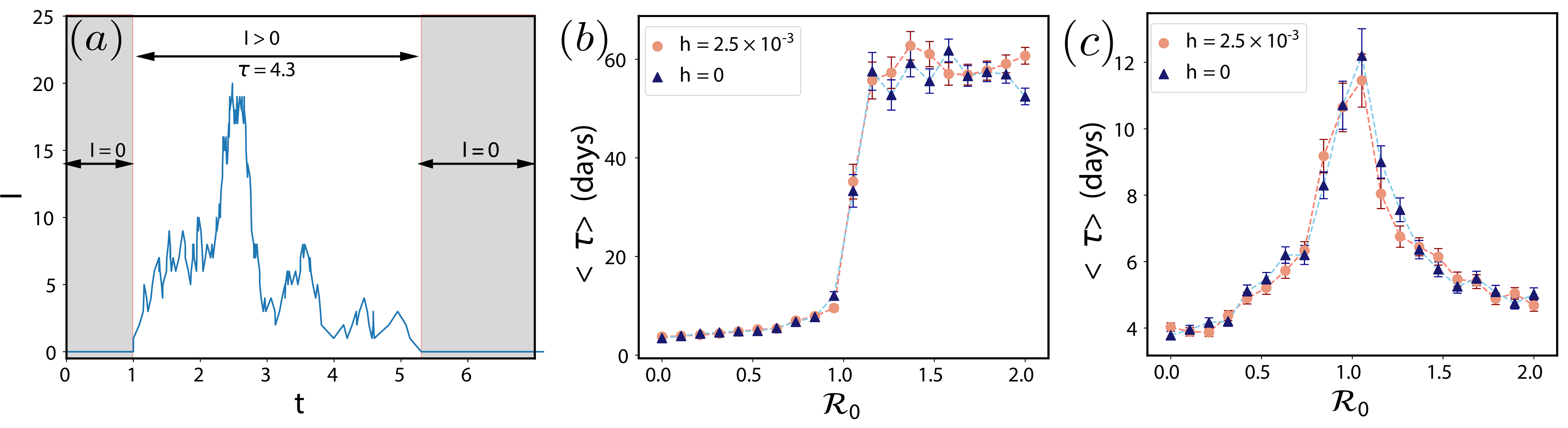}
 \caption{Results of numerical generation of trajectories of the stochastic SIR model with external seeding in a single population using the Gillespie algorithm\cite{gillespie1976general,toral2014stochastic}. (a) Example of a local outbreak of duration $\tau=4.3$ with parameters $N=10^5$ $\mathcal{R}_0=0.8$, $\mu=1/3.7\,\text{day}^{-1}$ and $h=1.2$. A shadowed area is placed where the disease is deactivated ($I=0$). (b) Average duration $\langle \tau \rangle$ of the first outbreak started from a single seed, $I(0)=1,\,S(0)=N,\,R(0)=0$ for different values of the external import ($h$) and of the basic reproductive number ($\mathcal{R}_0$). In (c), similar to (b), but the outbreak starts from an initial condition beyond herd immunity, using the initial conditions of  Eq.~\eqref{eq:IC_beyond_herd_immunity}. Both in (b) and (c), the results have been averaged over 100 realizations and the errorbars signal the magnitude of the standard error.}
 \label{fig:Local_outbreak} 
\end{figure*} 

 \subsection{With external seeding ($h>0$)}
 
If $h$ is non-zero, Eq.~(\ref{eq:SIRh_MF}) has only one 
fixed point irrespective of the values of $h$, $\beta$ and $\mu$:
\begin{align}
S_\infty\equiv&\lim_{t\rightarrow \infty}S(t)= \, 0 , \nonumber \\
I_\infty\equiv&\lim_{t\rightarrow \infty}I(t)= \, \frac{h}{\mu+\frac{h}{N}} , \label{eq:equil}\\
R_\infty\equiv&\lim_{t\rightarrow \infty} R(t)= \, N\frac{\mu}{\mu+\frac{h}{N}} .\nonumber
\end{align}
This fixed point is stable.
Therefore, there is no phase separation regarding the stationary state of the system. The external seeding removes the absorbing nature of the states with $I=0$ and the phase transition~\cite{marro2005nonequilibrium}. However, the dynamical evolution towards the fixed point will show differences depending on the values of the epidemic parameters. In order to see this, we investigate the behavior of Eq.~(\ref{eq:SIRh_MF}) with initial conditions: $I(0)=I_0\sim \mathcal{O}(1)$, $S(0)=N-I_0$ and $R(0)=0$ for a short time window and in the limit of large population, $N\gg 1$. In these limits, we find the linear approximation to Eq.~(\ref{eq:SIRh_MF}) for early times 
\begin{align}
\frac{dI}{dt} \approx \left( \beta-\mu-\frac{h}{N} \right)\, I + h ,
\end{align}
with solution 
\begin{align}\label{eq:ilineal}
 I(t) \approx I_0 \, e^{(\mathcal{R}^h_0-1)\, \mu \,t} +\frac{h/\mu}{\mathcal{R}^h_0-1}\left(e^{(\mathcal{R}^h_0-1) \,\mu t}-1\right),
\end{align}
where, for the sake of functional similarity, we have named the term $\mathcal{R}^h_0=\mathcal{R}_0- h /(N\, \mu)$ as the basic reproductive number in the presence of external seeding. If initial conditions without infected individuals are considered, $I_0=0$, then new outbreaks are still started by the external seeding. Although the equilibrium values given by Eq.~(\ref{eq:equil}) are independent of the value of $\mathcal{R}^h_0$, this parameter controls the characteristic time to reach the fixed point. For $\mathcal{R}^h_0>1$, the  number of infected individuals $I(t)$ will first increase exponentially and become of macroscopic order quickly, and then, due to the nonlinear terms in Eq.~\eqref{eq:SIRh_MF}, it will decrease towards the fixed point $I_\infty$. If $\mathcal{R}^h_0<1$, the evolution can be either monotonic or non-monotonic depending on the intensity of the seeding rate $h$, but in both cases the number of infected individuals will remain small through its entire evolution towards $I_\infty$. Therefore, for $\mathcal{R}^h_0<1$, the disease will still affect a macroscopic portion of the population but in a slow fashion. Interestingly, in the limit of small external seeding and big population size, in which we are interested ($h\sim \mathcal{O}(1)$, $N\gg 1$), the basic reproductive number for the SIR with or without external seeding are indistinguishable ($\mathcal{R}^h_0\approx \mathcal{R}_0$). Therefore, empirical methods to measure the basic reproductive number could not notice the presence of small external seeding. 

\subsection{Finite systems}

When stochastic effects are taken into account, the arrival of an infected individual triggers an epidemic outbreak during which the number of infected is different from zero. We will say that a population is active  when there is, at least, one infected individual, $I>0$. Contrary, an inactive population has $I=0$. The random duration $\tau$ of an outbreak is the time during which the population is active, (see Fig.~\ref{fig:Local_outbreak}(a) for a sketch).

The external seeding will create sequences of  consecutive outbreaks, as new infected individuals arrive at all times. If the average arrival time $h^{-1}$ is smaller that the average outbreak duration, $\langle \tau\rangle$, outbreaks are likely to overlap, while for $h^{-1}\gg \langle \tau\rangle$, the outbreak due to the arrival of an infected individual will not occur typically until a previous outbreak has disappeared. In the sub-critical regime $\mathcal{R}^h_0<1$, the outbreaks will be short, while in the case of $\mathcal{R}^h_0>1$, the first outbreak will likely generate a large number of infected individuals and, hence, its duration will increase significantly. In Fig.~\ref{fig:Local_outbreak}(b), it is shown that the average duration of the first outbreak $\langle \tau \rangle$ can be used to characterize the phase diagram of the single population SIR model with external seeding. By comparing with the line of $h = 0$, it shows evidence that a small external seeding does not produce a drastic change in the characteristic times of the dynamics.

Secondary outbreaks in the super-critical phase will, in general, be much smaller than the first one, both in intensity (number of infected individuals during the outbreak) and in duration, see Section~\ref{ap:effective_beta} of the Appendix. In Fig.~\ref{fig:Local_outbreak}(c) it is shown the average duration of the second outbreak after the first macroscopic wave. Instead of waiting until the first wave is over, we can force ``heard immunity" by starting the simulations from an initial condition:
\begin{equation}\label{eq:IC_beyond_herd_immunity}
    I(0)=1, \, R(0)=\alpha N,\,S(0)=N-I(0)-R(0),
\end{equation}
in which the fraction $\alpha$ of recovered equals the attack rate in the absence of external seeding, Eq.~\eqref{eq:attack_rate} with $s_0=1-1/N$. Also in Fig.~\ref{fig:Local_outbreak}(c), we show that the small external seeding does not alter the characteristic time of these outbreaks.

\begin{figure*}
 \centering
 \includegraphics[scale=0.4]{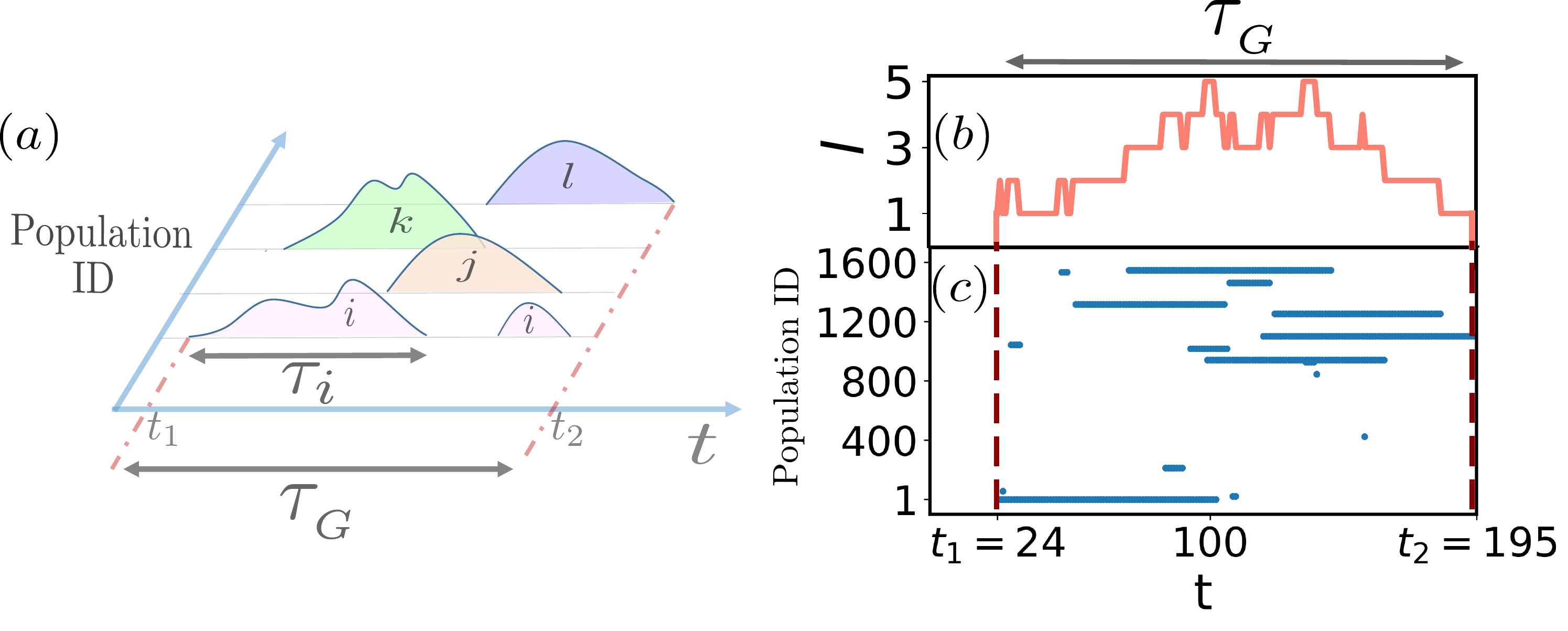}
 \caption{ In (a), we show a sketch with four different sub-populations, labeled $i$, $j$, $k$, and $l$, experiencing local outbreaks. Each of these outbreaks starts at a different time and has a different duration ($\tau$). They all contribute to a global outbreak of duration $\tau_{G}$. Such a global outbreak started with the first local outbreak in sub-population $i$ at time $t_1$ and finished at time $t_2$, when the last local outbreak died out (in sub-population $l$). In panels (b) and (c), we show the meaning of local and global outbreaks with actual simulations. In (b), we show the total number of infected individuals ($I=\sum_i I_i$) in a particular instance of a global outbreak. This global outbreak was initiated at time $t_1=24$, when the external seeding acted on the system with no other infected agent, and lasted until time $t_2=195$, when the total number of infected individuals became zero. The total duration of the global outbreak is $\tau_G=t_2-t_1=171$ days. In (c), using the same realization displayed in (b), we enquire about the duration of local outbreaks. The length of horizontal lines is the duration of local outbreaks, whereas the vertical axis informs about the label of the sub-populations. One can see how local outbreaks pile up generating the global outbreak of duration $\tau_G$. The parameters used to generate this example where $V=1600$, $\beta=0.8\mu$, and $h=0.1\, \text{days}^{-1}$ ($h/V=6.25\cdot10^{-5}$).}
 \label{fig:sketch}
\end{figure*}

\section{Independent sub-populations}\label{sec:independent_pop}


Throughout the rest of the work, we use a meta-population framework. This is, we deal with $V$ sub-populations, all of them having its own number of individuals (\{$N_i\}_{i=1,...,V}$) and separate compartment variables ($\{S_i,I_i,R_i\}_{i=1,...,V}$). Every sub-population follows a well-mixed stochastic SIR model. In the following, we fix the recovery rate to be compatible with the range of values for several important infectious diseases (such as COVID-19 or influenza): $\mu=1/3.7 \textrm{ days}^{-1}$~\cite{di2020impact}. By this means, we approach the time scales of real diseases and it is possible to grasp in a more intuitive way some results of this work, such as the order of magnitude of the outbreak survival times. The external seeding replaces an individual chosen at random between the whole system of sub-populations by a new infected individual. This is, a new infected individual enters the system at rate $h$, replacing an older individual chosen randomly within the $ \sum_i^V N_i$ total members of the whole population. Every sub-population $i$ is thus selected to receive the seed with a probability proportional to its population $N_i$.

We start with a simplistic setting in which all the sub-populations have the same number of individuals (namely, $N_i=10^5$, $\forall \, i \in [1,V]$) and are independent (there is no circulation of agents between them). In  this way, the external field is the only responsible for the onset of local epidemic outbreaks. This situation could model a strict lockdown in which mobility restrictions keep the sub-populations fully isolated. The external seeding is considered a small perturbation of such severe confinement. This simple approximation allows us to make analytical calculations and build the understanding of more realistic scenarios with communication between the sub-populations considered in the next section.


We are primarily interested in the anomalous epidemic fade-out after the first macroscopic wave of infections. For this reason and as in Eq.~\eqref{eq:IC_beyond_herd_immunity}, we fix the initial condition such that the total number of infected individuals is equal to zero and the fraction of recovered individuals is such that the possibility of macroscopic outbreaks is avoided. This means that for each sub-population $i$:
\begin{equation}\label{eq:IC_recovered}
    R_i(0)=\alpha \, N_i,\, I_i(0)=0, \, S_i(0)=N_i-I_i(0)-R_i(0).
\end{equation}
where $\alpha$ is the attack rate in the absence of external seeding (Eq.~\eqref{eq:attack_rate} with $s_0=1-1/N$). In this way, we mimic a situation in which the whole system suffered a major super-critical outbreak. 
This initial condition would be an absorbing state in the absence of external seeding ($h=0$), but its presence ($h>0$) opens the possibility to generate further outbreaks. We provide insights about the behavior of the system with different initial conditions in section~\ref{ap:Global_outbreak_first_wave} of the Appendix.


In the following, we will differentiate local from global properties. Being the local properties those referred to individual sub-populations, e.g., the prevalence in a sub-population $i$, $I_i(t)/N_i$, is local, while the total number of infected agents $I(t) = \sum_{i=1}^V I_i(t)$ is global. One of our objectives is to understand some global characteristics from the knowledge of the local ones. The main magnitude from which we will base the conclusions of this work is the duration of global outbreaks, $\tau_G$, defined as the time in which the number of infected individuals in the whole system remains strictly greater than zero.  If at time $t_1$ an external seed enters a system with no other infected individuals, $I(t_1)=1$, and the global prevalence remains different from zero until time $t_2>t_1$, this is, $I(t_2)=0$ but $I(t)>0$ for $t_1<t<t_2$, then the duration of a global outbreak would be $\tau_G=t_2-t_1$. In a more intuitive way, the uninterrupted concatenation of local outbreaks results in a global outbreak (see Fig.~\ref{fig:sketch}(a) for a sketch illustrating the difference between local and global outbreaks).

\subsection{Simulations}
The duration of global outbreaks is a random variable, and we will study its average value $\langle \tau_G \rangle$. Our first approach to examine the behavior of $\langle \tau_G \rangle$ is to compute it from direct simulations. Since we are interested in the anomalous
epidemic fade-out after the first macroscopic wave, our simulations will start from the initial condition defined in Eq.~\eqref{eq:IC_recovered}. Then, at some stochastic time $t_1$, the external seeding will generate one infected seed in sub-population label $i$. This event will start both a local outbreak in sub-population $i$ and a global outbreak. At a different time $t_2>t_1$ the total prevalence will be zero for first time after $t_1$. We will stop the simulation at $t_2$ and sample one value of the total duration of global outbreaks as $\tau_G=t_2-t_1$.  Let us remark that the local and global events differ since the external seeding could activate multiple sub-populations before $t_2$ (see Fig.~\ref{fig:sketch}(b) for an example of computation of $\tau_G$ in simulations). Repeating this experiment many times one can access the ensemble average $\langle\tau_G\rangle$. A technical difficulty arises since, as we will see, the values of $\tau_G$ can be prohibitively large in order to access them with simulations. Hence, in order to make affordable the computational cost of the work, in the simulations we set up an upper time limit $t_{\text{max}}$ after which we stop the simulation independently on whether the global outbreak has vanished or not. Therefore, the maximum value that one can sample for $\tau_G$ with this approach is $t_{\text{max}}$.
\begin{figure}
    \centering
    \includegraphics[scale=0.35]{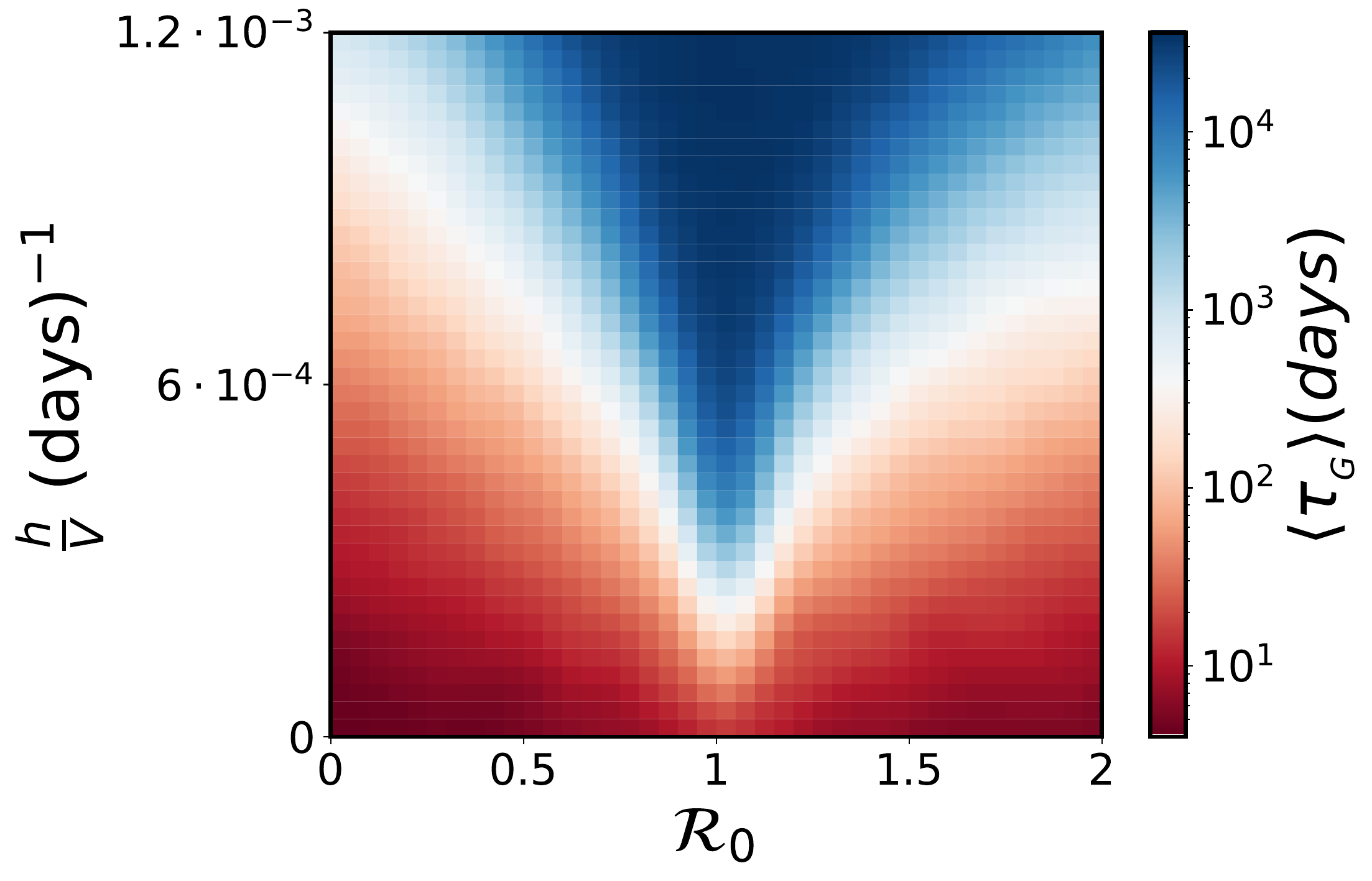}
    \caption{ In this figure, we investigate the average global outbreak duration, $\langle \tau_{G} \rangle$, for $V=1600$ independent (isolated) sub-populations obtained from numerical simulations using the Gillespie algorithm (\cite{gillespie1976general,toral2014stochastic}). Global outbreaks that do not end by the time $t_{\text{max}}=4\cdot10^4$ are stopped. The global outbreak duration was averaged over 100 realizations for different values of the basic reproductive number, $\mathcal{R}_0$, and the external seeding rate, $h$. A transition can be observed between a region where the duration of global outbreaks is of the order of local ones (in red, bottom left and right) and another one in which global outbreaks are much longer than local ones (in blue, center and top).}\label{fig:Global_outbreak_simu_M0}
\end{figure}
   
In Fig.~\ref{fig:Global_outbreak_simu_M0}, we plot $\langle \tau_{_G}\rangle$ as a function of the basic reproductive number $\mathcal{R}_0$ and the external seeding rate $h$. This figure informs thus about the time-scales for which the disease is active in the system. Interestingly, we can distinguish a cross-over between two regimes from the duration of global outbreaks: one in which the typical time-scale of global outbreaks is much bigger than the one of local outbreaks, and another in which the duration of global and local outbreaks share the order of magnitude. The presence of long global outbreaks after the first epidemic wave for small values of the seeding rate $h$ and for values of $\mathcal{R}_0$ both larger and smaller than one constitutes one of the main results of this paper. This provides a very simple mechanism to explain the observed endemic-yet-marginal epidemic states. 

There is an interplay between the external seeding and the epidemic dynamics. Given that the rate of activation ($h$) is small, if the duration of local outbreaks (tuned with $\mathcal{R}_0$) is too low, then the global outbreak is not sustained and will be quickly interrupted. Only if there is a proper balance between the two dynamics, we can observe the increase in the global outbreak duration (Fig.~\ref{fig:Global_outbreak_simu_M0}). Note that the single population setting cannot explain this effect. It is necessary the overlapping of sub-populations to generate the endemic state of the disease (Fig.~\ref{fig:Local_outbreak}(b)).


\begin{figure*}
 \centering
 \includegraphics[scale=0.4]{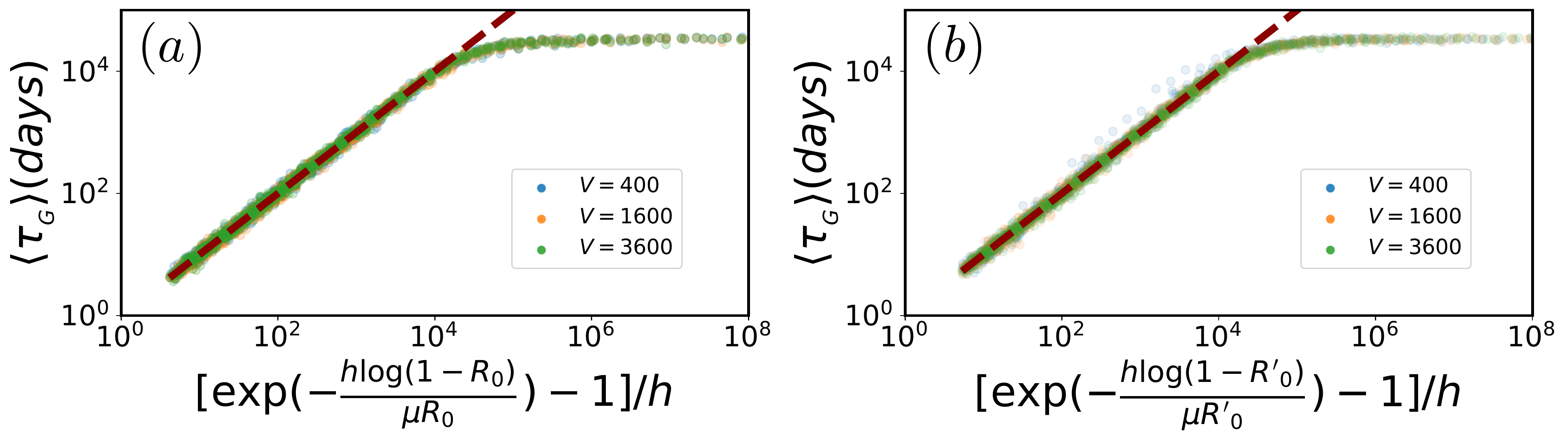}
 \caption{Comparison of analytical expressions for the average duration of global outbreaks with simulations. In (a), we show $\langle \tau_{_G} \rangle$ averaged over 100 realizations for different values of $V$, $\mathcal{R}_0$ and $h$. Here, we only use sub-critical values for the basic reproductive number ($\mathcal{R}_0<1$). The functional relation to collapse the data in a single curve is Eq.~(\ref{eq:Analitical_average_time_V2}), which is shown in dashed lines. In (b), we do the same, but concentrating on the super-critical region ($\mathcal{R}_0<1$), and using Eq.~\eqref{eq:Analitical_average_time_super_critical} to collapse the data. The dashed line is a plot of Eq.~\eqref{eq:Analitical_average_time_super_critical}. Both plots show evidence of the good agreement of the simulations with the theoretical predictions. The plateau observed in simulations is an artifact that  corresponds to the maximum time of simulations ($t_\text{max}$)} \label{fig:average_time_collapsed}
\end{figure*}

\subsection{Theory}


We define $n(t)$ as the number of active sub-populations (those for which the number of infected people is greater than zero) at time $t$. Our assumption is that we can write the evolution of the probability $P(n,t)$ that $n$ takes a certain value at a time $t$ in terms of a master equation with time-independent transition rates~\cite{van1992stochastic,gardiner1985handbook}. For this, we can write the variable $n(t)$ and the corresponding rates $W_+$ and $W_-$ of the master equation as follows
\begin{align}
W_+= & \lim_{dt\rightarrow 0} \frac{P(n(t+dt)=n_0+1|n(t)=n_0)}{(V-n_0) \, dt} , \\ 
 W_-=&\lim_{dt\rightarrow 0}\frac{P(n(t+dt)=n_0-1|n(t)=n_0) }{n_0 \, dt} . \nonumber
\end{align}
In this way, we find an expression for the duration of global outbreaks as a function of $W_-$ and $W_+$ in the limit of a large number of sub-populations, $V\gg 1$. To do so, we make use of the framework of the backward Kolmogorov equation to compute the average time to go from $n=1$ to $n=0$ (see~\cite{van1992stochastic,gardiner2009stochastic,krapivsky2010kinetic} and sections~\ref{ap:fixation_time} and~\ref{ap:two_state} of the appendix), which yields
\begin{align}
\label{eq:Analitical_average_time_V1}
\langle \tau_{_G} \rangle \sim \frac{1}{W_+ V} \, \left(e^{\frac{W_+}{W_-}\, V}-1\right).
\end{align}

In order to further exploit Eq.~(\ref{eq:Analitical_average_time_V1}), we need to identify the rates of the activation-deactivation process ($W_-$ and $W_+$) as functions of $\mathcal{R}_0$, $h$, $V$ and $N$. $W_+$ is the rate at which one inactive sub-population becomes active. Since the external seeding acts uniformly on every sub-population, we obtain that $W_+=h/V$. We associate $W_-$ with the inverse of the average time that a sub-population remains active: $W_-=1/\langle\tau\rangle$. In order to work with analytically tractable expressions, we approximate $\langle \tau \rangle$ for small $h$ with the average time with $h = 0$. When $h\approx 0$, it is unlikely that many external seeds enter in the same active sub-population; and even if so, they would not introduce big changes in the time scales (see Figs.~\ref{fig:Local_outbreak} (b) and (c)). Furthermore, we will treat differently the sub-critical ($\mathcal{R}_0<1$) and super-critical ($\mathcal{R}_0>1$) regimes. In the sub-critical region, we approximate the duration of local outbreaks by the average duration of a one-population outbreak in the SIS model (this statement is discussed in section  \ref{ap:fixation_time_SIR} of the appendix):
\begin{align}\label{eq:inverse_SIS_fixation_time}
W_- \approx -\dfrac{\mu \, \mathcal{R}_0}{\log(1-\mathcal{R}_0)} ,
\end{align}
see Appendix \ref{ap:SIS} for details on the derivation of Eq~\eqref{eq:inverse_SIS_fixation_time}. Therefore, we can rewrite Eq.~(\ref{eq:Analitical_average_time_V1}) as
\begin{align}
\label{eq:Analitical_average_time_V2}
\langle \tau_{_G} \rangle \sim \frac{1}{h}\, \left[\exp \left({\frac{-h\, \log(1-\mathcal{R}_0)}{\mu \, \mathcal{R}_0}}\right)-1\right].
\end{align}
This equation sheds light on the numerical results of Fig.~\ref{fig:Global_outbreak_simu_M0} for $\mathcal{R}_0<1$. In the first place, we can now reproduce the sub-critical region of this figure without an upper cut-off. Secondly, it allows us to collapse all the $h$, $\mathcal{R}_0$ and $V$ dependence in a single curve, see Fig.~\ref{fig:average_time_collapsed}(a). Moreover, it shows that the transition to large persistence times is not abrupt, the curve is continuous, non-divergent, and with well-behaved derivatives. Besides, it is of special interest that the scaling of $W_+$ with $V$ cancels the dependence of the average time with the global system size (see Fig.\ref{fig:average_time_collapsed}). Therefore, the general behavior of the long-lived epidemic states should not depend on the meta-population size and could be present at different scales (village, city or country level).


 In terms of the super-critical phase ($\mathcal{R}_0>1$), it is important to stress that sub-populations that reached local-herd immunity are still susceptible to generate further outbreaks induced by the external seeding $h$ or by infected visitors from other sub-populations. However, these outbreaks will not be macroscopic, since there is not a susceptible population large enough to fuel them. Our way to make quantitative predictions in this regime starts by noticing that the statistics of these outbreaks ``beyond herd-immunity" resemble those of the sub-critical regime in a susceptible population. Indeed, we can map the epidemic dynamics beyond herd-immunity by a sub-critical SIS model with a new effective transmission rate 
 \begin{equation}\label{eq:new_beta}
 \beta' = \beta \, (1-\alpha) ,
 \end{equation}
 where $\alpha$ is the attack rate defined in Eq.~\eqref{eq:attack_rate}. See section~\ref{ap:effective_beta} of the appendix for details on the derivation of Eq.~\eqref{eq:new_beta}. Therefore, we are able to exploit the same explanation built in for the sub-critical phase: even if local herd-immunity is reached, simultaneous sub-critical local outbreaks can percolate in time resulting in an endemic state at the global level. This observation enables us to estimate the average time of global outbreaks in the super-critical regime using an analogous version of Eq.~(\ref{eq:Analitical_average_time_V2}) with $\mathcal{R}'_0=\beta'/ \mu=-\mathcal{W}(-s_0\mathcal{R}_0e^{-\mathcal{R}_0})$, verifying $\mathcal{R}'_0\in(0,1)$ for $\mathcal{R}_0>1$ (section~\ref{ap:effective_beta} of the appendix):
 \begin{align}
\label{eq:Analitical_average_time_super_critical}
\langle \tau_{_G} \rangle \sim \frac{1}{h}\, \left[\exp\left({\frac{-h\, \log(1-\mathcal{R}'_0)}{\mu \, \mathcal{R}'_0}}\right)-1\right].
\end{align}

Eq.~\eqref{eq:Analitical_average_time_super_critical} predicts a similar collapse of data that the one observed in the subcritical regime but using now the effective transmission rate, see Fig.~\ref{fig:average_time_collapsed}(b).

\section{Adding mobility between sub-populations}\label{sec:mobility}

Although our results until now explain the emergence of epidemic endemic states, the assumption of independence between the sub-populations limits their applicability to real-world scenarios. A more realistic setting has to take into account that individuals can move across different sub-populations. This possibility enables a different mechanism to start local outbreaks, since infected agents can visit susceptible populations and susceptible individuals can also get infected out of their residence sub-population.

\subsection{Random diffusion}\label{subsec:diffusion}

Our first approximation to include mobility explicitly is pure random diffusion between sub-populations: every agent will jump to a connected neighboring sub-population at a constant rate $M$. For the moment, the number of connections per sub-populations is a constant ($k$), and the initial condition is uniform $N_i=N=10^5$ population across all sub-populations $i$. Under these circumstances, the distribution of the number of inhabitants will remain constant on average. Despite it has been shown that pure diffusion is not a proper description of human mobility in all scales, it has been used to model the large-scale spreading of infectious diseases (see, for example, the implementation of air traveling in ~\cite{keeling2000metapopulation,vespireview,colizza2007reaction, colizza2008epidemic,bajardi2011}). We shall see in the next section that the main results discussed here hold as well for the case of recurrent mobility.

Similarly to our procedure in section~\ref{sec:independent_pop}, we first investigate the duration of global outbreaks with direct simulations of the stochastic process in which, once more, we implement a maximum time $t_{\text{max}}$ at which simulations stop. As we are interested in the arising of anomalous outbreaks after the first wave, we will set in each sub-population the initial conditions given by Eq.~(\ref{eq:IC_recovered}). 

\begin{figure}
    \centering
    \includegraphics[scale=0.35]{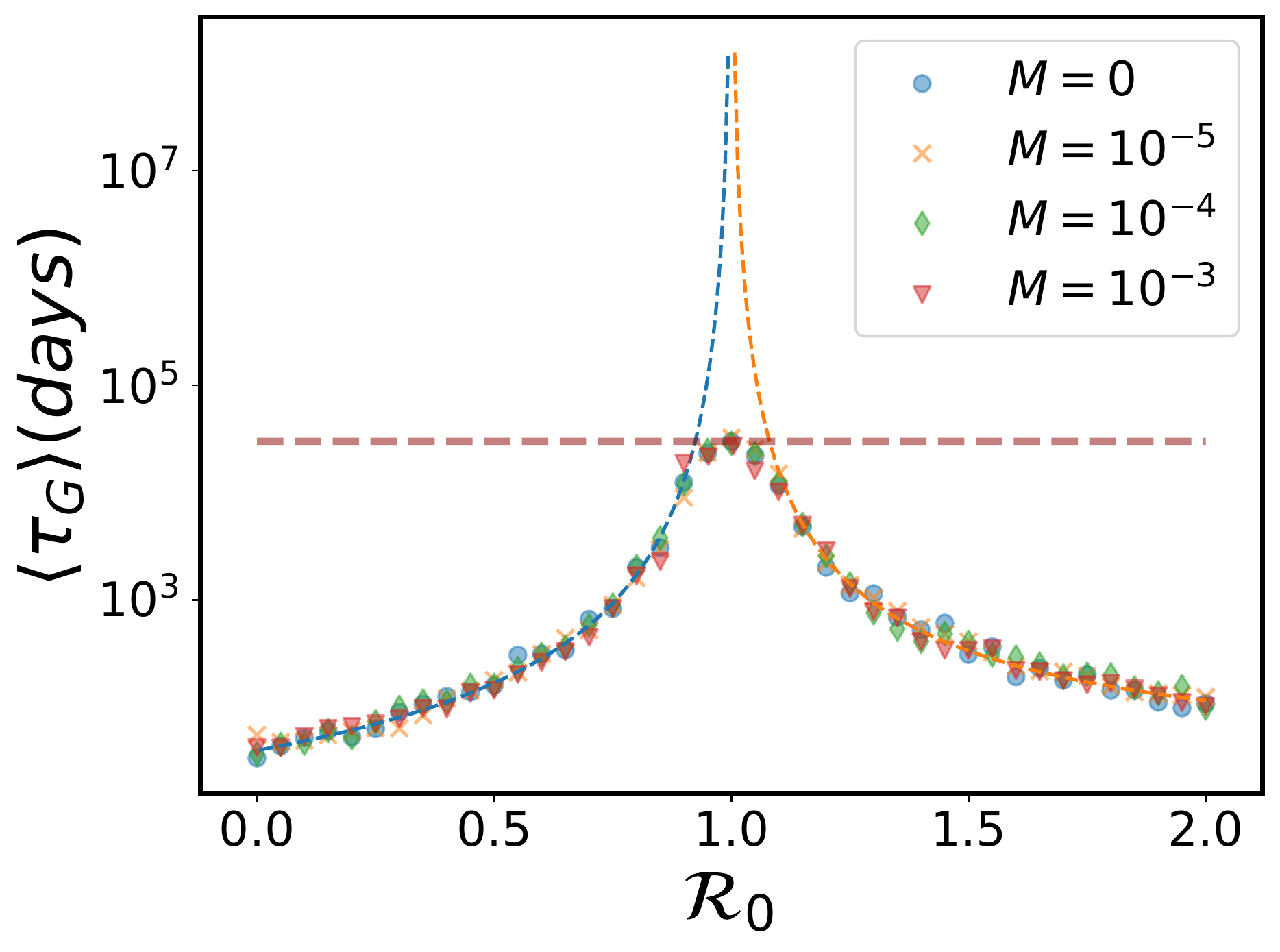}
    \caption{Effect of random diffusion on the duration of global outbreaks. Dots show the duration of the first global outbreak averaged over 100 simulations for different values of $\mathcal{R}_0$ and $M$. We show a transect of fixed external seeding ($h=1$ and $h/V=2.5\times 10^{-3} $). The initial condition mimics the situation after the first epidemic outbreak [Eqs.~\eqref{eq:IC_recovered}].  The topology is a squared lattice with periodic boundary conditions ($k=4$) with $V=400$\ ($20\times20$). All simulations are stopped either at time $t_{\text{max}}=4\times 10^4$ (days) (horizontal dotted line) or when the total prevalence reaches zero. Dashed curved lines show our analytical estimations (Eqs. (\ref{eq:Analitical_average_time_V2}) for $\mathcal{R}_0 < 1$ and (\ref{eq:Analitical_average_time_super_critical}) for $\mathcal{R}_0 > 1$). As discussed in the text, mobility doesn't have deep effect since no macroscopic outbreaks are expected.}\label{fig:Phase_diagram_mobility_diffusion_beyond_herd_immunity}
   \end{figure}
In Fig.~\ref{fig:Phase_diagram_mobility_diffusion_beyond_herd_immunity}, we show the average duration of the first global outbreak for different values of the mobility rate $M$ and the basic reproductive number $\mathcal{R}_0$. Essentially, the effect of mobility in the range of $M$ explored is negligible and the $\langle \tau_G \rangle$ of all simulations coincide with the theoretical prediction for the independent sub-populations case (Eq.~\eqref{eq:Analitical_average_time_V2} and Eq.~\eqref{eq:Analitical_average_time_super_critical}, both shown in dashed lines in Fig.~\eqref{fig:Phase_diagram_mobility_diffusion_beyond_herd_immunity}).

The probability that the first epidemic outbreak in a given sub-population $i$ affects a neighboring sub-population $j$ mainly depends on two factors: the number of infected individuals in $i$, and the rate $M$ at which individuals from $i$ travel to neighboring subpopulations~\cite{colizza2008epidemic}. This is the reason why in a situation with no macroscopic outbreaks, we do not expect mobility to play a major role. In Appendix~\ref{ap:Global_outbreak_first_wave}, we check that mobility does play a role in the duration of global outbreaks when considering the first wave in the analysis. In the same Appendix~\ref{ap:Global_outbreak_first_wave}, we also test that our results are robust to variations in the initial condition.

We note the predictive power of Eqs.~\eqref{eq:Analitical_average_time_V2} and~\eqref{eq:Analitical_average_time_super_critical} even in the case of mobility. Its applicability is remarkable, given the strong approximations introduced for its derivation (independent sub-populations and SIS dynamics).

\begin{figure}
\centering
\includegraphics[scale=0.2]{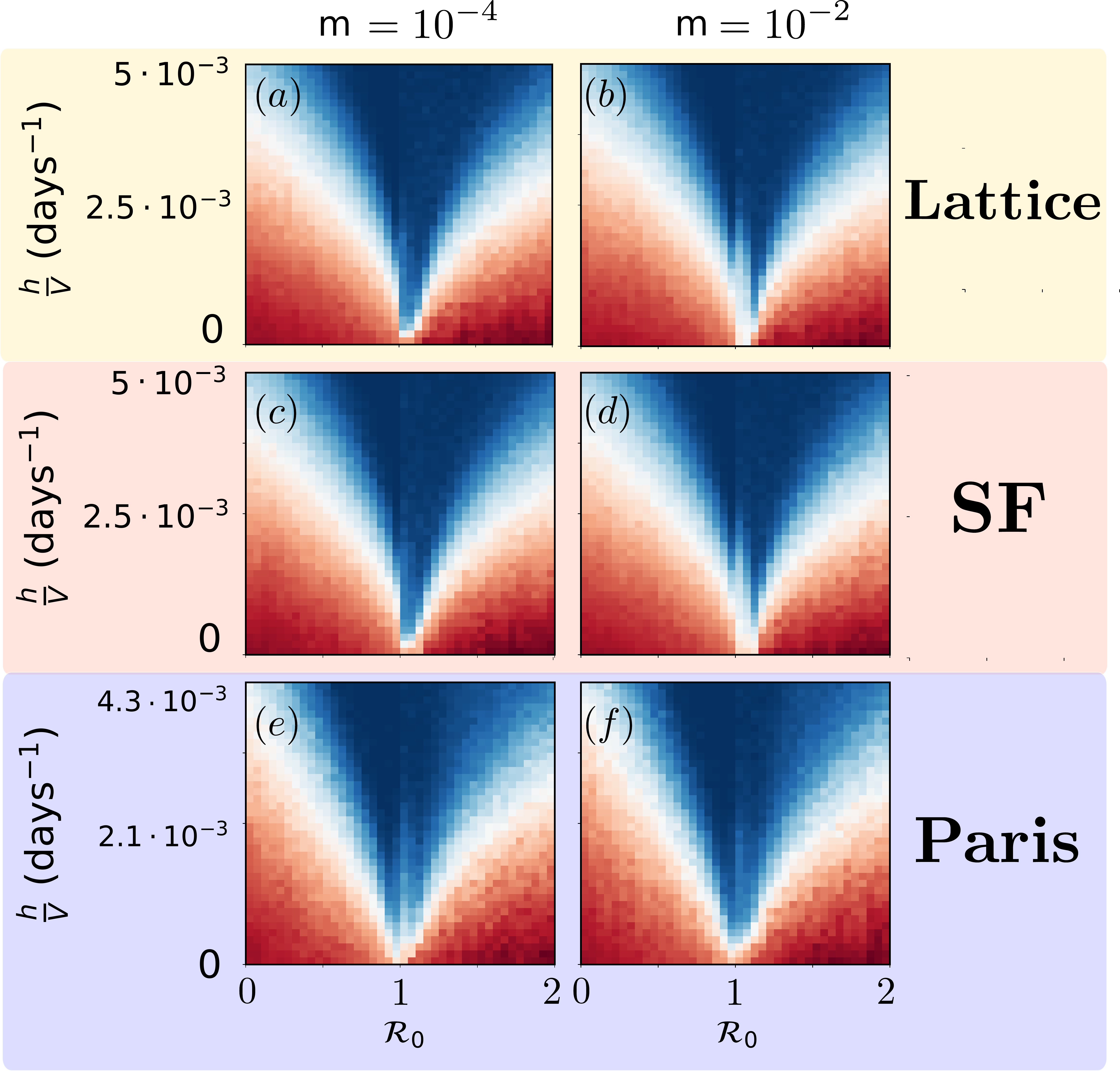} 
\caption{Average global outbreak duration for sub-populations connected through recurrent mobility, and for two values of the portion of travelers ($m$). Also,  different topologies and demographic statistics are inspected. Averages were performed over 100 realizations. Global outbreaks that do not end by the time $t_{\text{max}}=4\times 10^4$ are stopped.
In (a-b), $V=400$ sub-populations are connected forming a regular lattice with periodic boundary conditions. Demographics are Gaussian distributed. In (c-d), the topology is scale-free network with a degree distribution $P(k) \sim k^{-2.5}$ and with $V=400$ sub-populations proportional to the degree. In (e-f), connectivity and populations are read from commuting data of the city of Paris ($V=469$).}
\label{fig:Phase_diagram_t}
\end{figure}

\subsection{Recurrent mobility}\label{sec:recurrent_mobility}

We test next the robustness of our findings when the mobility is recurrent. This type of mobility is used to model back and forth trips as those related to commuting, which represent the majority of daily
mobility in urban environments. It has been proven that recurrent mobility produces different propagation patterns compared to diffusion due to the repetition of contacts in the residence and working areas~\cite{vespireview,balcan2011,arenas2020modeling,soriano2020,aguilar2020impact,mazzoli2021}. In practice, we assign to every agent a sub-population of residence and one of work (which can be the same). Agents are assumed to spend $1/3$ of the day in the working sub-population and the rest $2/3$ in the residence one. Note that this implies that the initial number of residents in each sub-population ($\{N_i\}_{i=1,...,V}$) is preserved in time. In this case, mobility fluxes are parameterized by the fraction of resident agents traveling every day ($m$). Once the daily fluxes between sub-populations are fixed, they remain the same during all the simulation. The main variables are thus the number of individuals living in sub-population $i$ and working in $j$ at each of the disease states ($\{X_{ij}\}$, where $i , j =1,...,V$, and $X$ can be $S,\, I \mbox{ or } R$). The addition of recurrent mobility makes it difficult to simulate the stochastic process using, for instance, the Gillespie algorithm in feasible times. In order to reduce the computing time, we make use of an approximation that exploits the difference between the time scales of the epidemic and mobility rates~\cite{balcan2009multiscale,balcan2010modeling}. The basic idea behind this approximation is that recurrent mobility is encoded in an effective transmission rate that depends on mobility and demographic characteristics of each sub-population.

We show next  that the regimes obtained in the previous sections still hold and they are not an artifact derived from the uniform distribution of populations and connections, nor of the specific type of mobility. In Fig. (\ref{fig:Phase_diagram_t}), we obtain similar patterns in the phase diagram as we vary the mobility intensity parameter ($m$) for:
\begin{itemize}
 \item (a-b) A configuration in which sub-populations form a 2-D regular lattice with a Gaussian distribution of the number of residents (average $10^5$ individuals and $\sigma = 3/20 \times 10^{5}$). The number of agents traveling in each link between sub-populations $i$ and $j$ are $m\, N_i/4$.
 \item (c-d) Sub-populations are connected by a scale-free network generated with the configurational model and with degree distribution $P(k) \sim k^{-2.5}$. The average degree is $\langle k \rangle = 6.2$. The outflow of a sub-population $i$ is equally distributed across the links departing from$i$ and it is equal to $m \, N_i/k_i$.
 \item (e-f) A realistic application in the city of Paris. The basic divisions of the city are census areas ``ensemble des communes", the resident populations and commuting networks are obtained from official statics~\cite{Demo_Paris,M_Paris}. As before, we use a control parameter ($m$) to determine the fraction of resident population that commutes. The destinations are selected according to the empirical commuting flows. For example, if $\omega_{ij}$ is the empirical number of individuals living in $i$ and working in $j$, we will consider in our simulations $m\, N_i \, \omega_{ij} / \sum_{\ell} \omega_{i\ell}$ travelers in the link $i-j$. 
\end{itemize}

\begin{figure}
\includegraphics[scale=0.4]{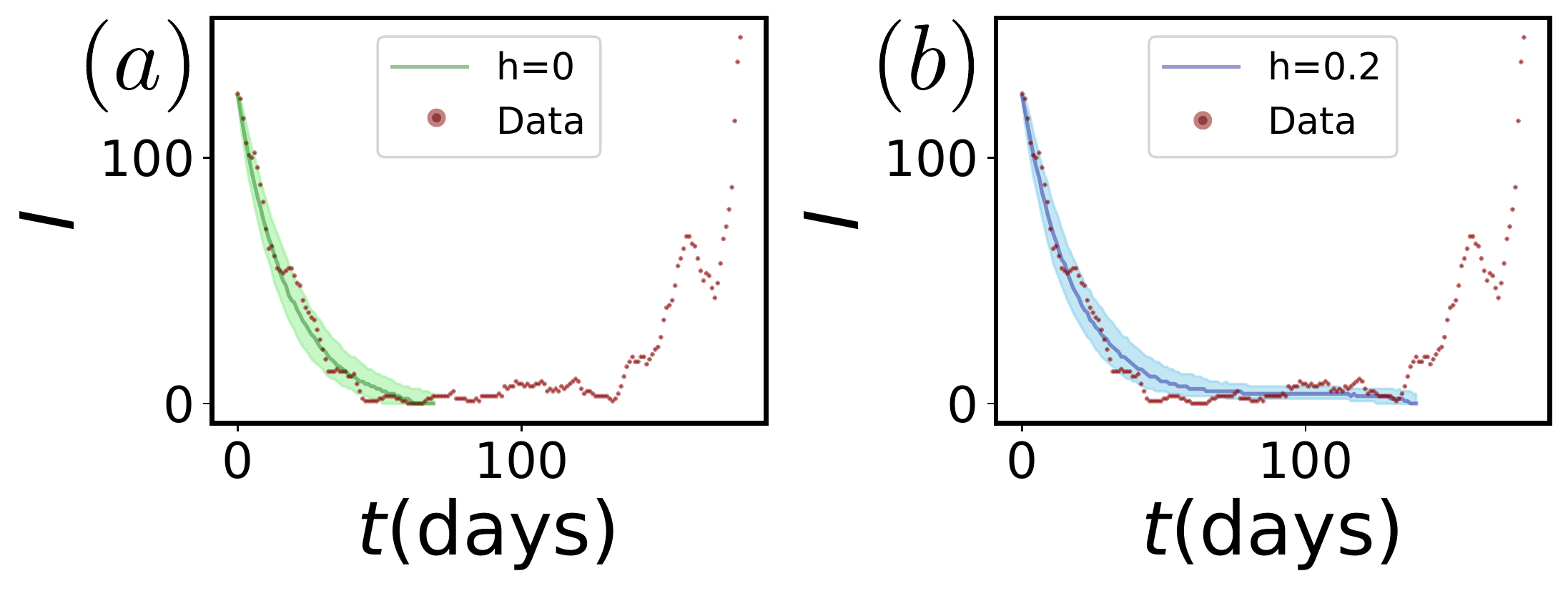}
 \caption{In the two panels we show with dots the evolution of the prevalence in one particular LTLA corresponding to the region of Haringey, in London. Observing the data, we differentiate three dynamical regions regarding the behavior of the prevalence corresponding to the first wave (exponential decay), the anomalous fade-out (fluctuating plateau), and the second wave (exponential growth). In (a) and (b), we also show results from simulations carried with the mobility switched off as described in section~\ref{sec:independent_pop}, using the demographic details of the LTLA and we fix $\mathcal{R}_0=0.8$. The solid line represents the median and the shadowed area of the first and third quartiles obtained from $10^3$ simulations. In (a), we show the evolution obtained with simulations without external seeding ($h=0$). With this setting, the model reproduces properly the exponential decay but fails in describing the subsequent plateau. In (b), simulations are run with $h=0.2$. In this case, the model captures both the decay and the plateau regimes.}
\label{fig:empirical_prevalence} 
\end{figure} 

In all these panels, we note that there is an extended parametric area for which the endemic states emerge. These results are robust to changes in the initial condition (see section~\ref{ap:Global_outbreak_first_wave} in the appendix) and to the epidemic model: i.e., a SEIR model also generates the same variety of behaviors (see section \ref{ap:SEIR} in the appendix).

\section{Empirical evidence} \label{sec:empirical}

Lastly, we compare our predictions with publicly available governmental data of COVID-19 spreading  
in England~\cite{incidence_govUK}. Specifically, we focus on the anomalous epidemic fade-outs observed in COVID-19 incidence (number of new infections per day) between the two first waves of the pandemic. A situation that exactly represents the assumptions of our model. Broadly speaking, this period corresponds to the months between April and September 2020 and the geographical resolution of our data is at the level of the ``lower-tier local authorities" (LTLA) --one of the administrative units in which the country is divided. England is composed of 315 (see a sketch in Fig.~ \ref{AP-fig:LTLA_map}) of these divisions whose average population is $179,000$ inhabitants.

\begin{figure}[b]
 \includegraphics[scale=0.3]{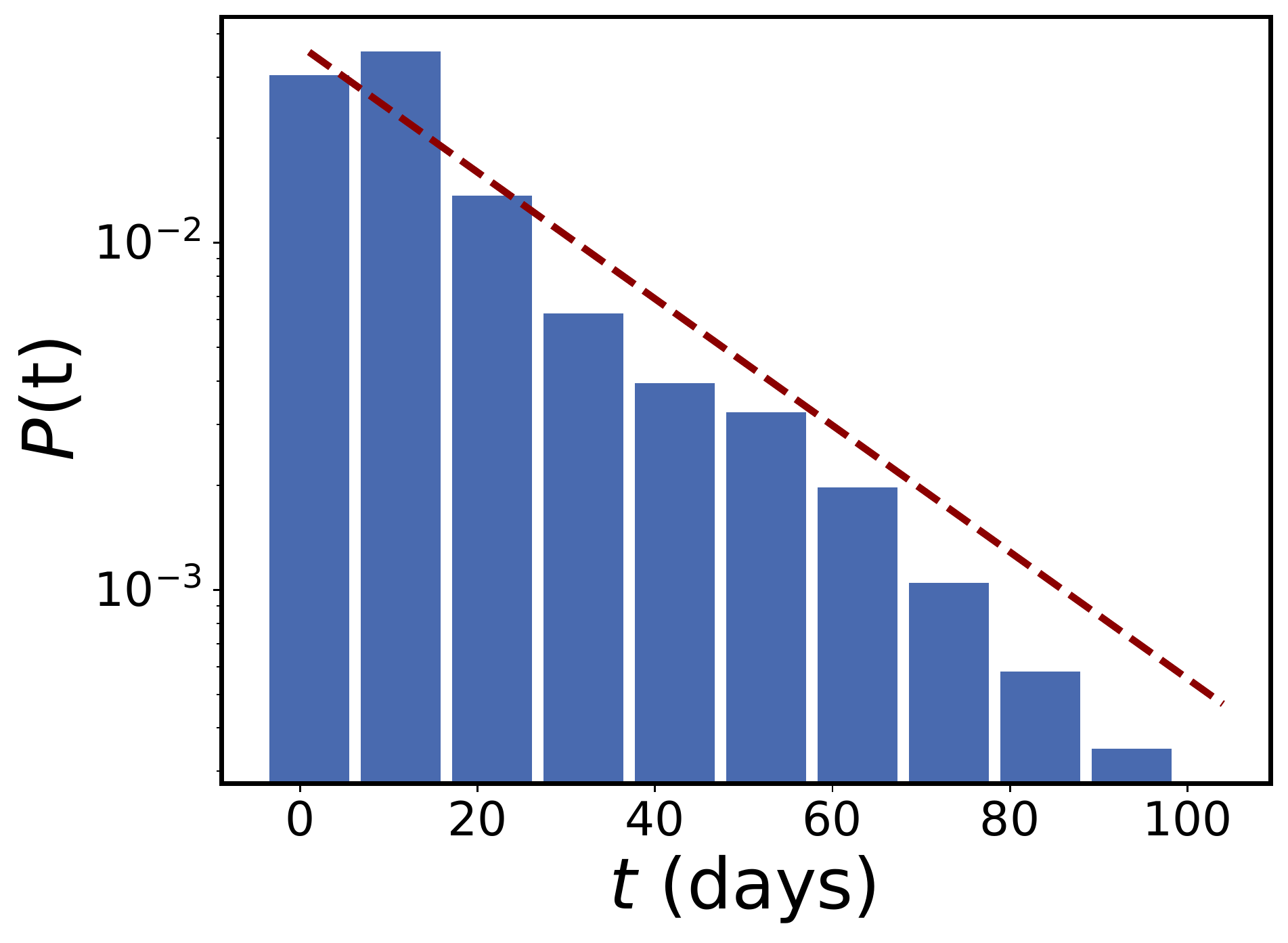}
\caption{Distribution of times for which the LTLAs have zero prevalence, as predicted by the theory, is well fitted by an exponential distribution (Eq.~\eqref{eq:exponential_distribution}). The value of the exponent of the best fit is $0.041(8)$, and can be used as a proxy for the rate at which infected individuals enter the LTLAs from outside per unit of time.}
\label{fig:empirical_inter_event_time} 
\end{figure} 

\begin{figure}
\includegraphics[scale=0.3]{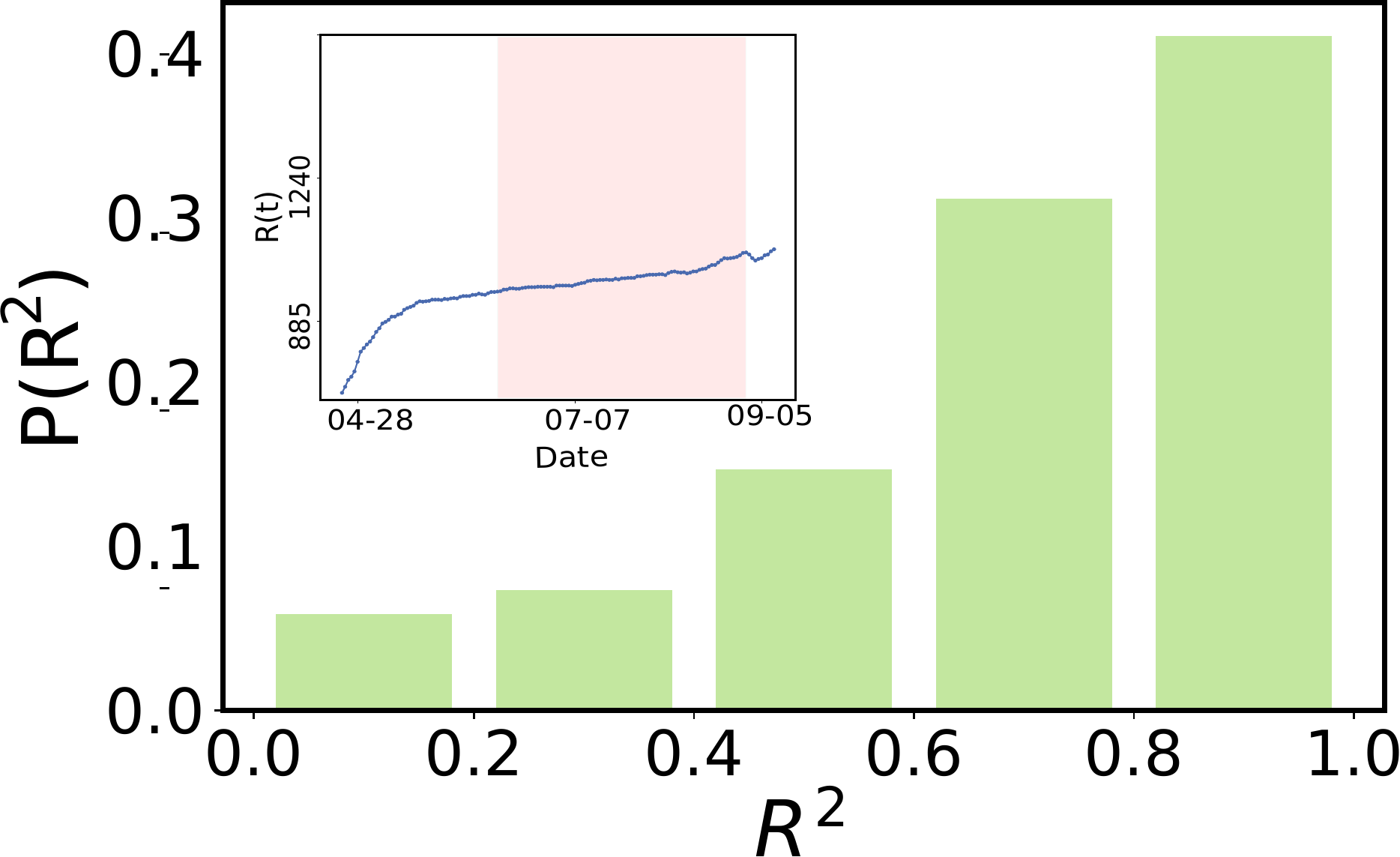}
 \caption{For each LTLA we fit the evolution of the number of recovered individuals to a linear function during the period between the first two epidemic waves. In the figure, it is shown the probability distribution of the coefficients of determination ($R^2$) resulting from the linear fit. The number of recoveries is measured as the cumulative of the incidence minus the prevalence ($R(t) = \sum_{t'=0}^{t'=t} \text{inc}(t')  - I(t')$). The evolution of the number of recoveries is well-fitted by a linear function in the majority of LTLAS ($R^2\approx 1$). The linear growth is the one expected by our model for any external seed when the prevalence is close to zero and $\mathcal{R}_0<1$ (see Fig.~\ref{AP-fig:examples_linear_simulations}). However, this kind of linear growth would only be present at the critical point of the classic SIR model without external field. Those LTLAs in which the number of recoveries do not follow a linear function ($R^2\approx 0$) are associated with LTLAs in which the activity between waves was null or very low (see Appendix~\ref{AP-fig:examples_linear}). The inplot shows one example of evolution of the number of recoveries together with the shadow area signaling the period in which the curve is well-fitted by a linear function. More examples are provided in the Appendix~\ref{AP-fig:examples_linear}.}
\label{fig:empirical_linear} 
\end{figure} 
Disease prevalence (number of infected individuals per day) is computed from the empirical incidence by assigning to each new case an incubation and an infectious period. The first one is sampled from a log-normal distribution, with a mean incubation period of $5.2$ days, parameterized as in~\cite{Li2020}. The infectious period follows a exponential distribution with mean $2.3$ days chosen as in~\cite{Di_Domenico2020}. Since incidence data is only weekly available, we uniformly distribute cases over the days of the week to facilitate the analysis and comparison with the simulation results. In Fig.~\ref{fig:empirical_prevalence}, we show one instance of the evolution of the prevalence in a particular LTLA. In the data, we can differentiate two regimes with exponential changes in the prevalence, associated with the decay and growth of the first and second wave respectively.  We can also see a third dynamical phase between the two  waves, in which prevalence flattens and is low but almost always non-zero. This phase is what we call an anomalous persistent fade-out since it cannot be characterized by the standard models (see Fig.~\ref{fig:empirical_prevalence}(a)). However, as shown in Fig.~\ref{fig:empirical_prevalence}(b), our model equipped with an external field is able to reproduce both the exponential decay and the subsequent fluctuating plateau.

Fig.~\ref{fig:empirical_inter_event_time}  adds more quantitative information to our discussion. It shows that the distribution of times for which LTLAs have zero prevalence is well-fitted by an exponential functional form. This is precisely the distribution expected by our model when neglecting mobility of infected individuals. In this case, the activation of an LTLA can only be caused by the field and the distribution of times with zero prevalence would read
\begin{equation}\label{eq:exponential_distribution}
    P(t)=h\, e^{-h\, t}.
\end{equation}
Where $h$ is the rate at which infected individuals enter the LTLAs from outside. Hence, one can estimate the external field from the exponential fit of the distribution in Fig.~\ref{fig:empirical_inter_event_time} together with Eq.~\ref{eq:exponential_distribution}, in this case, obtaining $h\approx 0.04$ (days)$^{-1}$ as a proxy for the external seeding rate at LTLA level. 

\begin{figure}
\includegraphics[scale=0.35]{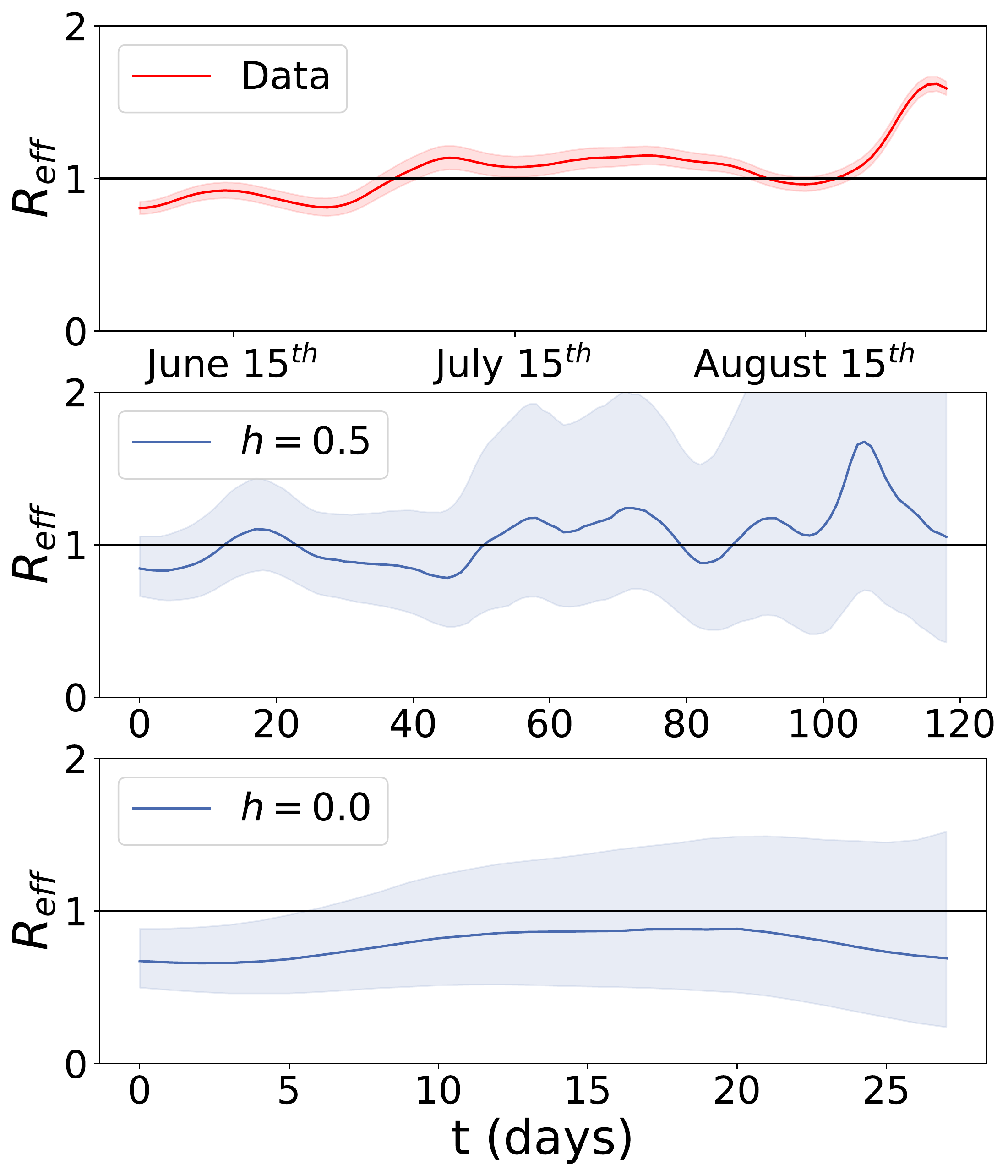}
\caption{Measures of the effective reproductive number. All measures shown of the time-varying reproduction number are computed with the method explained in~\cite{Cori2013} and its associated package. In (a), we show the effective reproductive number estimated from the incidence in England during the period between the first and second Covid-19 waves. In (b) and (c), we show measures of the basic reproductive number over one realization of our model as described in section~\ref{sec:independent_pop} with $V=32$, $N=8000$, thus representing the average size and population of an LTLA. In both (b) and (c), the initial condition is $I(t=0)=185$, $S(0)=V*N-185$ and $\mathcal{R}_0=0.8$, thus resembling the epidemic state of of the LTLA shown in Fig.~\ref{fig:empirical_prevalence} on March 2020 as measured from the real data. In (b), there is an external field ($h=0.5$), and we recover the near-critical and fluctuating values for the effective reproductive number observed in real data [this is, in (a)]. In (c), the external seeding is switched-off ($h=0$), and the values of the effective reproductive number are more monotonous and clearly sub-critical.}
\label{fig:R0_effectivo} 
\end{figure}

With the previous results we have shown that our model is capable of reproducing the statistics of epidemics in the period within the two first waves. Furthermore, external seeding can also explain 
features of anomalous fade-outs that resemble those of fine-tuned critical points. Indeed, in Fig.~\ref{fig:empirical_linear}, we show that the growth of recovered individuals during the period within the two first waves is well-fitted by a linear function. This linear growth is a general characteristic in our model that we would expect for any external field with $I\approx0$ and $\mathcal{R}_0<1$, however, the linear growth is only shown at the critical point in the standard SIR model~\cite{Radicchi2020}. A different signature of criticality is  given by the  measures of the effective  reproductive number obtained in real data, which fluctuate around the critical value (see Fig.~\ref{fig:R0_effectivo}(a)). Remarkably, we can measure similar kind of fluctuating and near-critical values for the effective reproductive number on data generated with simulations of a metapopulation SIR model with external seeding (see Fig.~\ref{fig:R0_effectivo}(b)). In Fig.~\ref{fig:R0_effectivo}(c), we show that this behavior disappears when the external seeding is switched-off. In this case, we observe more monotonous and clearly sub-critical values for the basic reproductive number. In Figs. \ref{AP-fig:h0}, \ref{AP-fig:h05} and \ref{AP-fig:h1} of the Appendix we show that this phenomenon is observed in a robust way for different values of the external seeding.

\section{Conclusions} \label{sec:conclusions}

We have proposed and studied the addition of a small external field to a SIR dynamics on a meta-population system. This field accounts for the important rate of infectious or latent individuals undetected to the surveillance systems and who can arrive from other populations or even reside in the local one. We show that small external fields are not noticeable by usual estimates of the basic reproductive number, yet they can have notorious effects at the global scale. Our findings are general and not restricted to a specific disease. However, they are specially well-suited for the COVID-19 situation, in which non-vaccinated regions could act as reservoirs of undetected infections at low-yet-constant rates.

Our main result is that a small external seeding can cause epidemic endemic states for an extended parametric region. This phenomenon has relevant consequences: 1) Even if the pharmaceutical and non-pharmaceutical response to an epidemic crisis can ensure that the transmissibility becomes sub-critical, it cannot be granted that the disease fades out. The spreading survives in a low-prevalence, yet uninterrupted epidemic state. The danger of these persistent states is that the system is highly susceptible to generating new exponential outbreaks as soon as control measures are lifted or new variants emerge. 2) For super-critical scenarios, we also show that herd immunity in all sub-populations does not imply an extinction of epidemics at the global level. This fact echoes the results of~\cite{keeling2000metapopulation}, which showed that rats acting as a reservoir of bubonic plague remove the concept of herd immunity even if the full population is vaccinated. In our case, it is not necessary a reservoir species since humans from other populations by themselves act as the reservoirs. Thus, we join a recent current of works claiming that the whole notion of herd immunity must be revisited~\cite{aschwanden2021five,tkachenko2021time}. 

The framework of the backward Kolmogorov equations, used to compute fixation times, allowed us to check our numerical findings analytically and obtain scaling relations. Moreover, it shows that this phenomenon is not linked with a sharp transition around a tipping point. The map of the SIR model to a two-state system conceptually means a coarse-graining of the local dynamics. This strategy could be further exploited in the future in order to deal with the local-global complex relation inherent to any meta-population structure.

This work is specially pertinent as the current literature is struggling to find explanations to criticality signatures found in the COVID-19 spread (uninterrupted-yet-small prevalence, linear growth of the recoveries, high susceptibility to changes in mobility restrictions and social distancing, etc). Our model is capable of reproducing the persistence of the COVID-19 disease between
waves in the census areas of England. We can also explain empirical features such as  the exponential distribution of the time between outbreaks, the linear growth of the recoveries and the
near-critical values of the effective reproductive number. These results are remarkable given the simplicity of our assumptions and the lack of fine-tuning. Our model is not equipped with the explicit time-dependence needed to capture the arising of new macroscopic
prevalence peaks (that we link to reduction of the restriction measures and the arising of new variants). Although it is possible to develop a multi-strain version, we kept the model simple in this work for the sake of  analytical tractability.


\section{Data Availability statement}

 Resident populations and commuting networks of the city of Paris are obtained from the official statics department~\cite{Demo_Paris,M_Paris}. Epidemic data on COVID-19 is obtained from UK government sources~\cite{incidence_govUK}.

\section{Code Availability statement}

The codes for the different models and data analysis are available at~\cite{Github} and are free to use providing the right credit to the author is given.

\bibliography{refs}

\begin{acknowledgments}
We thank Aleix Bassolas for his help in data manipulation. Partial financial support has been received from the Agencia Estatal de Investigaci\'on and Fondo Europeo de Desarrollo Regional (FEDER, UE) under project APASOS (PID2021-122256NB-C21/PID2021-122256NB-C22), and the María de Maeztu project CEX2021-001164-M, funded by the  MCIN/AEI/10.13039/501100011033. 
\end{acknowledgments}

\section*{Author Contributions Statement}  

J.A., B.A.G., R.T., S.M. and J.J.R. conceived and designed the study. B.A.G collected the epidemic data. J.A. and B.A.G analyzed the data. J.A., R.T., S.M. and J.J.R. developed and adapted the models. J.A. performed the simulations. J.A., B.A.G., R.T., S.M. and J.J.R. wrote the paper. All the authors read and approved the paper. 

\section*{Competing Interests Statement}

The authors declare no competing interests.

\clearpage
\newpage
\onecolumngrid
 \clearpage

 \setcounter{page}{1}

 \begin{center}
 \Large{SUPPLEMENTARY MATERIAL\\ Endemic infectious states below the epidemic threshold and beyond herd immunity\\ }
\end{center}
\begin{center}
 Javier Aguilar, Beatriz Arregui Garc\'ia, Ra\'ul Toral, Sandro Meloni and Jos\'e J. Ramasco, \\
Instituto de F\'{\i}sica Interdisciplinar y Sistemas Complejos IFISC (CSIC-UIB), Campus UIB, 07122 Palma de Mallorca, Spain.
\end{center}
\hrulefill
\setcounter{figure}{0}
\setcounter{equation}{0}
\setcounter{section}{0}
\renewcommand{\thesection}{S\arabic{section}} 
\renewcommand{\theequation}{S\arabic{equation}}
\renewcommand{\thefigure}{S\arabic{figure}}
\setcounter{equation}{0}

\section{Attack rate in the SIR model}\label{ap:attack_rate_SIR}

In this section, we derive the analytical form of the attack rate for the SIR deterministic model in the absence of external seeding. The attack rate is defined as the total number of agents affected by the disease,
\begin{equation}
    \alpha := \lim_{t\rightarrow \infty} \frac{R(t)}{N}.
\end{equation}
We focus on the deterministic case,
\begin{align}
    \label{eq:SIR_MF_V1}
     \frac{dS}{dt}&=-\beta \, I\, \frac{S}{N}, \nonumber \\
     \frac{dI}{dt}&= \beta \, I \, \frac{S}{N} - \mu\, I , \\
     \frac{dR}{dt}&= \mu \, I. \nonumber
\end{align}
The basic strategy is to re-parametrize Eqs.\eqref{eq:SIR_MF_V1} in order to get rid of the time in the description of the process:
\begin{equation}\label{eq:I_as_function_of_s_SIR_MF}
    \frac{dx}{ds}=-1+\frac{1}{\mathcal{R}_0 s}
\end{equation}
Where $x$ and $s$ are the density of infected and susceptible individuals respectively, $x=I/N$ and $s=S/N$; and $\mathcal{R}_0=\beta/\mu$. We can solve Eq.~\eqref{eq:I_as_function_of_s_SIR_MF} integrating from the initial fraction of susceptible individuals, $s_0=S(t=0)/N$ to the final point $s_f=S(t=t_f)/N$:
\begin{equation}\label{eq:exact_solution_SIR}
    x(s_f)-x(s_0)=-(s_f-s_0)+\mathcal{R}_0^{-1}\log(\frac{s_f}{s_0}).
\end{equation}
We now invert the above equation in order to express the final fraction of susceptible individuals as a function of $x_f=x(s_f)=I(t=t_f)/N$, $x_0=x(s_0)=I(t=0)/N$, and $\mathcal{R}_0$.
Where $\mathcal{W}$ is the Lambert function, which fulfills $\mathcal{W}(z)\exp(\mathcal{W}(z))=z$.
\begin{equation}
    s_f=-\mathcal{R}_0^{-1}\mathcal{W}\left(-\mathcal{R}_0e^{\mathcal{R}_0(x_f-x_0-s_0+\mathcal{R}_o^{-1}\log({s_0}))}\right).
\end{equation}
Setting $x_0+s_0=1$ and $x_f=0$ (no recovered individuals at initial time and no infected individuals at final time), one can compute the attack rate as:
\begin{equation}\label{eq:ap_attack_rate}
    \alpha =1-s_f=1+\mathcal{R}_0^{-1}\mathcal{W}\left(-s_0\mathcal{R}_0e^{-\mathcal{R}_0}\right).
\end{equation}

\section{Average fixation time for a two-state process} \label{ap:fixation_time}
We derive here the analytical expression of the average fixation times using the backwards Kolmogorov equation (also called adjoint equations). Although this procedure is well known in the literature (e.g.~\cite{krapivsky2010kinetic,gardiner2009stochastic,van1992stochastic}), and seeking for self-containment, we summarize key results of this framework. In doing so, we will follow closely the exposition by~\cite{krapivsky2010kinetic}. Then we will detail our strategy to solve such equation in some systems of interests. 

We focus on two-state processes, meaning that agents can be in either one of two possible states ($\sigma_i=0$ or $\sigma_i=1$ for $i=1,\dots,N$). The macroscopic state of such systems is characterize by just one occupation number $n$, which we define as the number of agents in state $1$:
\begin{equation}
    n=\sum_{i=1}^{N}\sigma_i.
\end{equation}
In all, we will treat an stochastic process with discrete states ($n$) and continuous time ($t$). If we discretize time in bins of width $dt$, the average time $T(n)$ to go from a state with $n$ agents in state $1$ towards a state with $0$ agents in that state satisfies:
\begin{eqnarray}
 T(n)=dt+\sum_{\ell=-1,0,1}P(n(t+dt)=n+\ell|n(t)=n)T(n+\ell)+O(dt^2),
\end{eqnarray}
where the $O(dt^2)$ accounts for multiple processes within the time interval $(t,t+dt)$ and we assume that the only possible processes are those that alter $n\to n\pm1$. Dividing by $dt$ and introducing the transition rates:
\begin{eqnarray}
 & \lambda(n) := \lim_{dt\rightarrow 0} \frac{P(n(t+dt)=n+1|n(t)=n)}{dt} \nonumber \\
 & \gamma(n) := \lim_{dt\rightarrow 0} \frac{P(n(t+dt)=n-1|n(t)=n)}{dt}
\end{eqnarray}
we obtain, after taking the limit $dt\to 0$, the difference equation:
\begin{eqnarray}
 \label{eq:map_average_fixation_time}
 -1=-\left[\lambda(n)+\gamma(n)\right]T(n)+\lambda(n)T(n+1)+\gamma(n)T(n-1)
\end{eqnarray}

We will solve the above equation with boundary conditions $T(0)=0$ (absorbing) and $T(N-1)-T(N)=-\frac{1}{\gamma(N)}$ (reflecting). In all the cases of study, the annihilation rate fulfills $\gamma(N) \propto N$. Since we are interested in the thermodynamic limit $N\gg 1$, we can therefore approximate $T(N-1)-T(N)=0$. Eq.~(\ref{eq:map_average_fixation_time}) can be solved using a change of variables:
\begin{eqnarray}
 \label{eq:def_U}
 U_n:=T(n)-T(n+1) \rightarrow T(n)=-\sum_{i=0}^{i=n-1} U_n.
\end{eqnarray}
Where the boundary condition $T(0)=0$ was used. Also note that the reflecting boundary condition implies $U_{N-1}=0$. Eq.~(\ref{eq:map_average_fixation_time}) in terms of $U_n$ reads:

\begin{eqnarray}
 U_n=\frac{\gamma(n)}{\lambda(n)}U_{n-1}+\frac{1}{\lambda(n)}.
\end{eqnarray}

The homogeneous part of the above equation,
\begin{eqnarray}
 U_n^H=\frac{\gamma(n)}{\lambda(n)}U_{n-1}^H,
\end{eqnarray}
is easy to solve, $U^H_n=U^H_0 \prod_{i=1}^n \frac{\gamma(i)}{\lambda(i)}$. This result suggests the use of the ansatz $U_n=V_n R_n$, where
\begin{eqnarray}
 R_i:=
 \begin{cases}
      \prod_{j=1}^{j=i} \frac{\gamma(j)}{\lambda(j)}, & \text{if}\ i>0 \\
      1, & \text{if}\ i=0
\end{cases}.
\end{eqnarray}

The new map in terms of $V_n$ reads:
\begin{eqnarray}
 V_n=V_{n+1}-\frac{1}{\lambda(n) R_n}.
\end{eqnarray}
We just have to iterate above map to find its solution:
\begin{eqnarray}
 V_n=V_{N}-\sum_{i=n}^{N-1}\frac{1}{\lambda(i) R_i}.
\end{eqnarray}
We now invert the two changes of variables and use the reflecting boundary condition to find :

\begin{eqnarray}
 T(n)=\sum_{j=1}^{n}R_{j-1}\sum^{N-1}_{i=j}\frac{1}{\lambda(i)R_i}.
\end{eqnarray}

We are particularly interested in computing the time to go from state $n=1$ to state $n=0$:
\begin{eqnarray}
 \label{eq:time_1}
 T(1)=\sum_{i=1}^{N}\frac{1}{\lambda(i)R_i}.
\end{eqnarray}
where we have replaced $N-1$ by $N$ as the upper limit of the sum as we are interested in the limit $N\rightarrow\infty$.

\section{Fixation time in independent two-state systems.}\label{ap:two_state}

 Say a system can be in one out of two possible states : ``activated" ($\sigma=1$) or ``deactivated" ($\sigma=0$). We define $W_+$ the transition rate to go from $0$ to $1$ and $W_-$ the rate to go from $1$ to $0$. A collection of $V$ independent copies of such system is characterized by the number of activated states:
 \begin{eqnarray}
 n(t):=\sum_i^V \sigma_i(t).
 \end{eqnarray}
 We want to compute the average time to go from $n=1$ to $n=0$. To do so, we make use of Eq.~(\ref{eq:time_1}). Firstly, we identify the transition rates that define the process $n(t)$.
 \begin{eqnarray}
 \label{eq:rates_two_state_pop}
 & \gamma(n)=n W_-, \nonumber \\
 & \lambda(n)=(V-n)W_+.
 \end{eqnarray}
 Then, we compute the auxiliary variables $R_i$:
 \begin{eqnarray}
 \label{eq:Ri_two_state_pop}
 R_i=\prod_{j=1}^{j=i} \frac{\gamma(j)}{\lambda(j)}=\left(\frac{W_-}{W_+}\right)^i \frac{i!}{\prod_{j=1}^{j=i}(V-j)}.
 \end{eqnarray}
 Defining $r=\frac{W_+}{W_-}$ and inserting Eqs. (\ref{eq:rates_two_state_pop}) and (\ref{eq:Ri_two_state_pop}) in Eq.~(\ref{eq:time_1}):
 \begin{eqnarray}
 \label{eq:T1_two_state_no_approx}
 T(1)=\frac{1}{W_+}\sum_{i=1}^V \frac{r^i}{i!}\prod_{j=1}^{j=i-1}(V-j)=\frac{1}{VW_+}\sum_{i=1}^V r^i \frac{V!}{i!(V-i)!}=\frac{1}{V W_+} [(1+r)^V-1].
 \end{eqnarray}
 Since we are interested in cases in which $W_+\sim \mathcal{O}(V^{-1})$ and $W_-\sim \mathcal{O}(1)$, we can approximate:
 \begin{eqnarray}
 \label{eq:T1_two_state_with_approx}
 T(1)\approx \frac{1}{V W_+}(e^{V r}-1). 
 \end{eqnarray}

\section{Fixation time in SIS model.} \label{ap:SIS}
In this section, we compute the average duration of an outbreak originated from a single seed in the Susceptible-Infected-Susceptible (SIS) model. We focus in the thermodynamic limit ($N=I+S\gg 1$). We make use of Eq.~(\ref{eq:time_1}), to do so, we first identify the transition rates that define the process $I(t)$.
 \begin{eqnarray}
 \label{eq:rates_SIS}
 & \gamma(I)=\mu I \nonumber \\
 & \lambda(I)=\beta I \frac{N-I}{N},
 \end{eqnarray}
 where $\beta$ and $\mu$ are the infection and recovery parameters respectively. Then, we compute the auxiliary variables $R_i$:
 \begin{eqnarray}
 \label{eq:Ri_SIS}
 R_i=\prod_{j=1}^{j=i} \frac{\gamma(j)}{\lambda(j)}=\left(\frac{N\mu}{\beta}\right)^i \frac{1}{\prod_{j=1}^{j=i}(N-j)}.
 \end{eqnarray}

 Recalling $\mathcal{R}_0=\frac{\beta}{\mu}$ and inserting Eqs. (\ref{eq:rates_SIS}) and (\ref{eq:Ri_SIS}) in Eq.~(\ref{eq:time_1}):
 \begin{eqnarray}
 \label{eq:T1_SIS_no_approx}
 T(1)=\frac{1}{\beta}\sum_{i=1}^N\frac{\mathcal{R}_0^i}{i N^{i-1}} \prod_{j=1}^{i-1}(N-j)
 \end{eqnarray}
 In the sub-critical phase $\mathcal{R}_0<1$ and the terms of the above series converge to zero. The product makes it difficult to find for a closed form. In the limit $N\to \infty$ We can approximate $\prod_{j=1}^{i-1}(N-j) \approx N^{i-1}$ and obtain:
 \begin{eqnarray}
 \label{eq:T1_SIS_with_approx}
 T(1)\approx \frac{1}{\mu \mathcal{R}_0}\sum_{i=1}^\infty \frac{\mathcal{R}_0^i}{i}=-\frac{1}{\mu \mathcal{R}_0} \log(1-\mathcal{R}_0) 
 \end{eqnarray}

 \begin{figure*}
 \centering
 \includegraphics[scale=0.35]{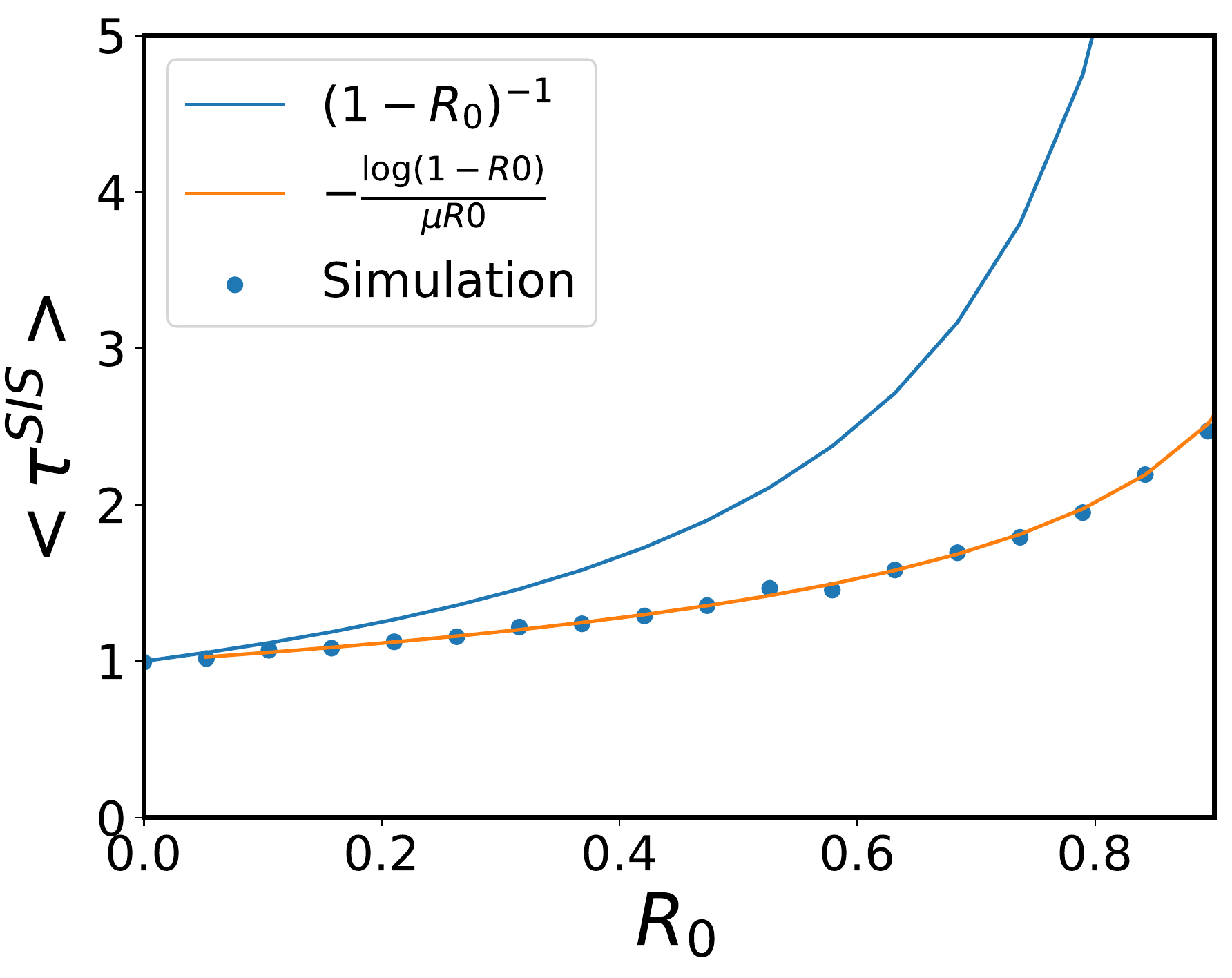}
 \caption{Average extinction time from a single seed in the SIS model. The analytical result given by Eq.~(\ref{eq:T1_SIS_with_approx}) is plotted together with the results from simulations with $N=10^6$ (dots). The behavior obtained with the deterministic mean-field approximation, this is, inverting Eq.~\eqref{eq:ilineal}  to obtain $T\sim \frac{1}{1-\mathcal{R}_0}$, is also shown.}
 \label{fig:SIS_time}
\end{figure*}

Comparisons of our analytical findings with numerical calculations is shown in Fig. \ref{fig:SIS_time}.
 
 \section{Fixation time for a one-step process} \label{ap:time_one_step}
 In this section, we estimate the average time that it takes for the collection of independent populations to reach a state of homogeneous local herd immunity. Say that at rate $\Tilde{W}_+$ deactivated populations are infected with a seed that generates a macroscopic outbreak. We also need to define now $n$ as the number of populations that reached herd immunity. Since local herd immunity is irreversible, we can rewrite Eq.~(\ref{eq:map_average_fixation_time}) for a single jump process.

 \begin{eqnarray}
 -1=-\Tilde{W}_+ T_{n}+\Tilde{W}_+T_{n+1}.
 \end{eqnarray}
 
 This equation can be solved with an absorbing boundary at $V$, i.e $T_V=0$, to obtain the average time to go from $n=0$ to $n=V$:
 \begin{eqnarray}
 T_0=\frac{V}{\tilde{W}_+}.
 \end{eqnarray}
We can approximate $\tilde{W}_+$ by the rate at which a seed enters in one population times the probability that this seed generates offspring before getting recovered. This is, $\tilde{W}_+=\dfrac{h}{V}\dfrac{1}{\mathcal{R}_0^{-1}+1}$ . Therefore:
 
 \begin{eqnarray}
 T_0=\frac{\mathcal{R}_0^{-1}+1}{h}V^2.
 \end{eqnarray}
 

We note that this simple estimation incorrectly assumes that populations obtain herd immunity instantaneously. However, since the time to reach herd immunity is a local property, we don't expect that a proper consideration of this fact would change the scaling of $T_0$ with $V$.
 
\section{Fixation time in the SIR model}\label{ap:fixation_time_SIR}
In the SIR model, agents can be in three possible states. Since the total number of agents is conserved ($N$), the macroscopic state of the system can be described by two occupation numbers, that we choose to be $I$ and $S$ (total number of infected and susceptible individuals respectively). Therefore, we have to modify the derivation of section~\ref{ap:fixation_time} that concerns only processes in which agents can have two states (and therefore can be described with only one occupation number). The reasoning to write the equation for the average fixation time $T(I,S)$ to reach the absorbing state $I=0$ starting from a state $(I,S)$ is similar to that followed in section~\ref{ap:fixation_time}.
\begin{equation}\label{eq:fixation_time_three_state_process}
\begin{split}
        T(I,S)=dt&+P(I,S;t+dt|I,S;t)T(I,S)\\
        &+P(I+1,S-1;t+dt|I,S;t)T(I+1,S-1)\\
        &+P(I-1,S;t+dt|I,S;t)T(I-1,S)+O(dt^2).
\end{split}
\end{equation}

For the particular case of the SIR model, we  remember the form of the transition rates:
\begin{eqnarray}
     & \beta I \frac{S}{N} = \lim_{dt\rightarrow 0} \frac{P(I+1,S-1;t+dt|I,S;t)}{dt}, \nonumber \\
     & \mu I = \lim_{dt\rightarrow 0} \frac{P(I-1,S;t+dt|I,S;t))}{dt}.
\end{eqnarray}
Inserting the above equations in Eq.~\eqref{eq:fixation_time_three_state_process}, taking the limit $dt\to 0$, using $\mathcal{R}_0=\frac{\beta}{\mu}$, and rearranging terms we finally obtain the recursion equation:
\begin{equation}\label{eq:solution_fixation_time_SIR}    
        T(I,S)=\frac{\mu^{-1}N}{I(N+\mathcal{R}_0S)}+\frac{T(I-1,S)N}{N+\mathcal{R}_0 S}+\frac{T(I+1,S-1)\mathcal{R}_0 S}{N+\mathcal{R}_0 S}.
\end{equation}

We have not been able to find the solution of Eq.~\eqref{eq:solution_fixation_time_SIR} in a closed form. However, starting from the boundary values $T(0,S)=0$ for $S=0,\dots,N$, one obtains $T(I,0)$ iterating Eq.~\eqref{eq:solution_fixation_time_SIR} for $I=1,\dots,N$. Recursively one then computes $T(I,1)$, for $I=1,\dots,N-1$ (note that $I+S\le N$), then $T(I,2)$, for $I=1,\dots,N-2$, and so on up to $T(1,N-1)$. This procedure can be implemented numerically allowing us to compute all fixation times with the only errors coming from the numerical accuracy of the computer arithmetic.

\subsection{Sub-critical approximation}
A key point of our analytical procedure was to approximate the fixation times in the sub-critical phase of the SIR by the corresponding fixation times in the SIS process. The great advantage of this approximation is that we can replace the non-closed expression for the SIR Eq.~\eqref{eq:solution_fixation_time_SIR} by the closed form in the case of the SIS model [Eq.~\eqref{eq:T1_SIS_with_approx}]. In the sub-critical phase, the outbreaks will not be of macroscopic order. As the number of infected individuals (and thus, the number of recoveries) will remain low, one would expect that the recovered individuals will not play a major role in the duration of the outbreaks and that this approximation is reasonable.
\begin{figure*}
\centering
\includegraphics[scale=0.4]{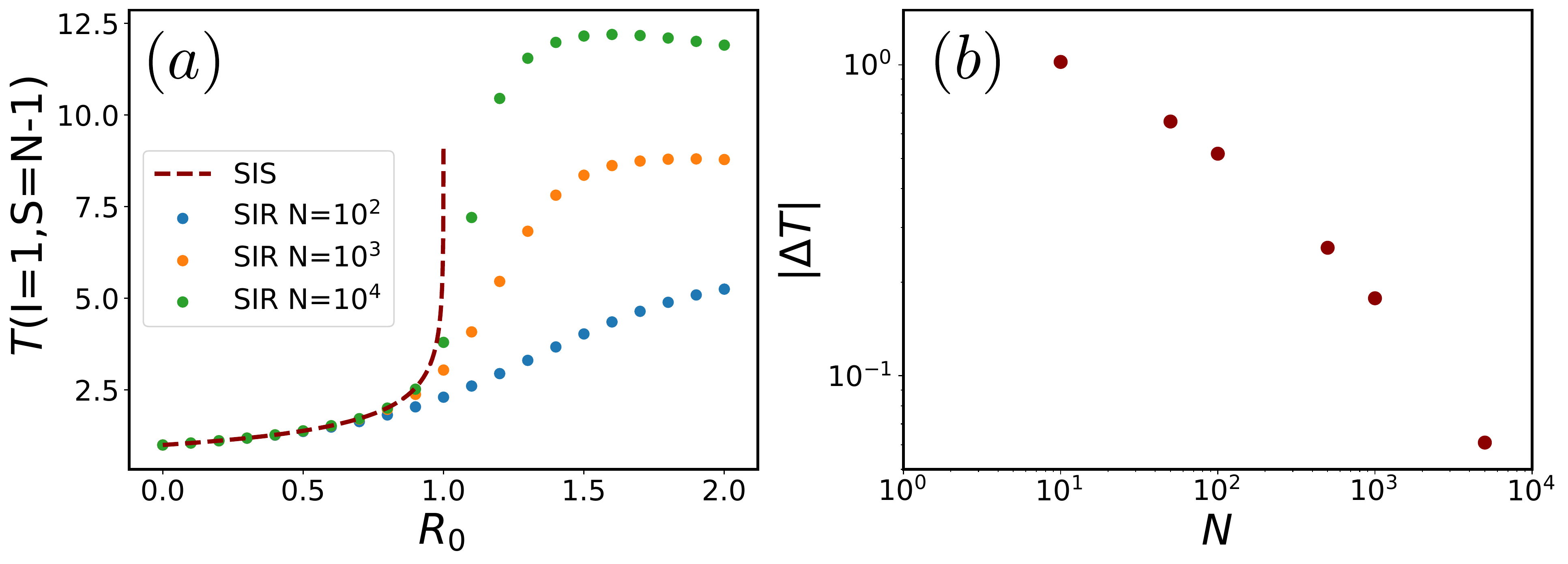}
\caption{ In (a), we show the average time to go from $(I=1,S=N-1)$ to the absorbing state ($I=0$). In dashed line, we plot Eq.~\eqref{eq:T1_SIS_with_approx}, which is our closed form for the absorption time in the SIS process bellow the critical point $\mathcal{R}_0=1$ (sub-critical regime). The points correspond to the solution for the case of the SIR process iterated numerically [Eq.~\eqref{eq:solution_fixation_time_SIR}]. It can be observed that both expressions are in good agreement in the sub-critical phase, except for the behavior in a neighborhood of $\mathcal{R}_0=1^-$. In particular,the average time of outbreaks is divergent at $\mathcal{R}_0=1$ in the case of the SIS, whereas our expression for the SIR process gives finite times at the critical point. In (b), we show the absolute value of the difference of the absorption times with the SIS and SIR models for different sizes $N$ and fixed $\mathcal{R}_0=0.9$. It can be observed that the difference between the two models decreases as $N$ increases.}
\label{fig:outbreaks_beyond_herd_immunity}
\end{figure*}
We note that it is difficult to bound the systematic errors induced by this approximation.  The reason is that the procedure is not a perturbation method that makes use of a small parameter.  We rather propose an alternative model (the SIS) in which we can advance with the numerical treatment of closed expressions. However, we can compare the results found with the SIS model [Eq.~\eqref{eq:T1_SIS_with_approx}] with those of the SIR model [Eq.~\eqref{eq:solution_fixation_time_SIR}]. In Fig.~\ref{fig:outbreaks_beyond_herd_immunity}-(a) we provide such a comparison, finding that both expressions are generally in good agreement except in the vicinity of the critical point $\mathcal{R}_0=1$, where the discrepancy has its origins in the finite-size effects. Note that in the case of the SIS model, we introduced the limit $N\to\infty$ and there is no $N$-dependence in Eq.~\eqref{eq:T1_SIS_with_approx} whereas, in the exact treatment of Eq.~\eqref{eq:solution_fixation_time_SIR} we are able to obtain the dependence in $N$. In Fig.~\ref{fig:outbreaks_beyond_herd_immunity}-(b), it is explicitly shown how the discrepancies between Eq.~\eqref{eq:T1_SIS_with_approx} and Eq.~\eqref{eq:solution_fixation_time_SIR} get reduced as $N\rightarrow\infty$.

\section{Effective transmission rate $\beta'$}\label{ap:effective_beta}
 
Say that we study a single population with $N\gg 1$ individuals equipped with SIR dynamics. A few infected individuals in an otherwise healthy population in the super-critical phase ($\mathcal{R}_0>1$) will, in general, trigger an outbreak of macroscopic dimensions ($I_{max}\propto N$).

\begin{figure*}
    \centering
    \includegraphics[scale=0.4]{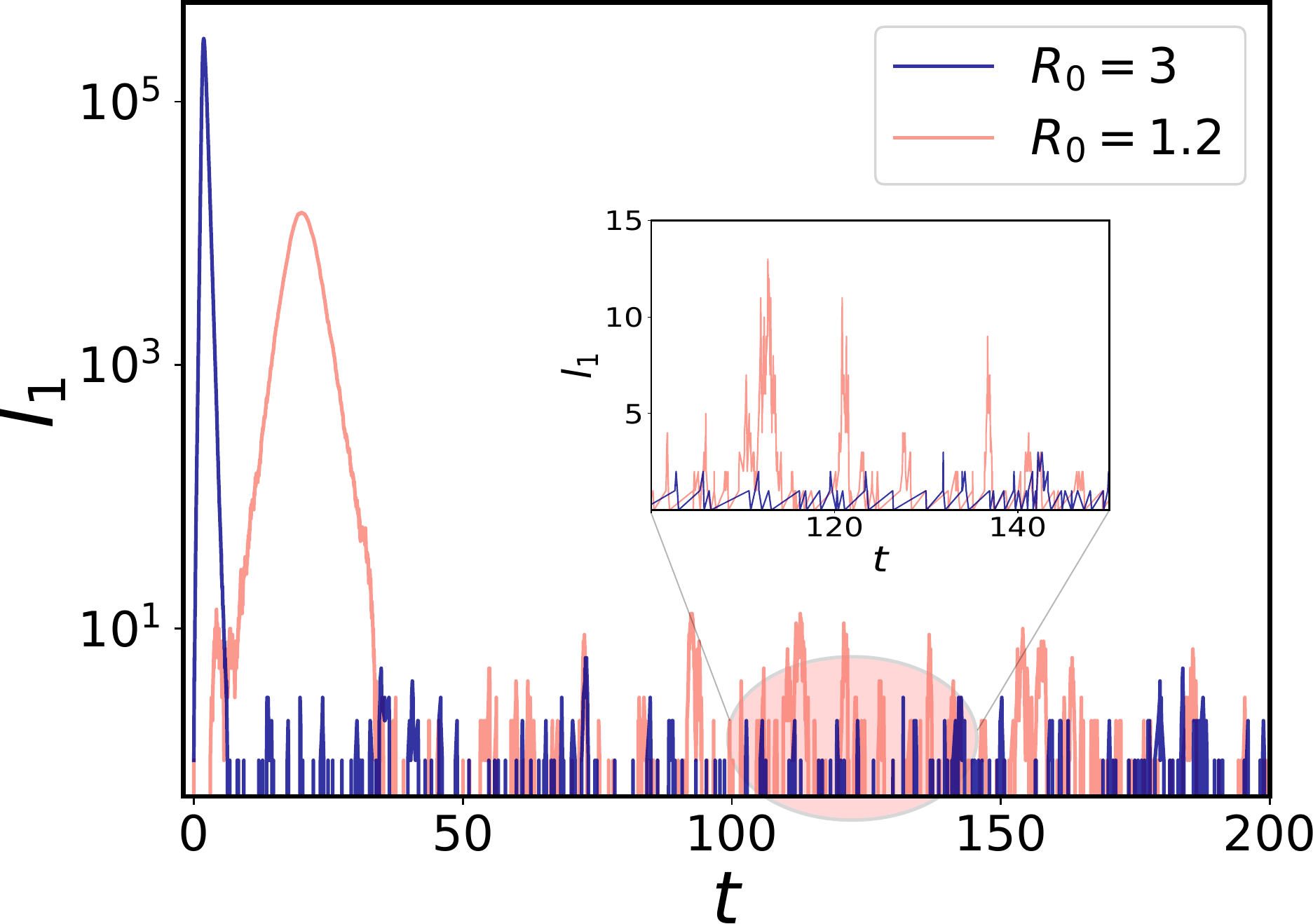}
    \caption{The prevalence for two realizations of the super-critical SIR process with different values of $\mathcal{R}_0$. The first peak in both curves is of macroscopic proportions. The subsequent peaks of the outbreaks (once herd immunity is reached) are activated by an external seeding and will not scale with the system size. The inset plot offers a zoom of the curves in linear scale. Notice that outbreaks beyond herd immunity are, in general, more pronounced for lower values of $\mathcal{R}_0$.}
\label{fig:outbreaks_beyond_herd_immunity_realizations}
\end{figure*}

If after this outbreak we perturb the absorbing state with infected seeds, subsequent outbreaks will no longer scale with the size of the system (as shown in Fig. (\ref{fig:outbreaks_beyond_herd_immunity_realizations})). This is another way to state the herd-immunity property at local level. Using the mean-field equations:
\begin{align}\label{eq:SIR_MF}
 \frac{dS}{dt}&=-\beta I \frac{S}{N}, \nonumber \\
 \frac{I}{dt}&= \beta I \frac{S}{N} - \mu I, \\
 \frac{R}{dt}&= \mu I,\nonumber
\end{align}
we could describe this non-macroscopic outbreaks using the initial conditions $S(0)/N=N(1-\alpha)-1$, $I(0)=1$ and $R(0)=N \alpha$, where $\alpha$ is the attack rate as defined in Eq.~\eqref{eq:attack_rate} with $s_0=1-1/N$. The attack rate is representative of the average size of the first macroscopic outbreak. As the number of susceptible individuals monotonically decreases, its possible values are bounded to the range $[0,N(1-\alpha)]$. We now propose the change of variables 
\begin{eqnarray}
 S'=S\frac{1}{1-\alpha}.
\end{eqnarray}
Note that the effective range of $S'$ now is $[0,N]$ and the deterministic equations read:
\begin{align}
 \frac{dS'}{dt}&=-\beta I \frac{S'}{N} (1-\alpha) \nonumber \\
 \frac{dI}{dt}&= \beta I \frac{S'}{N} (1-\alpha) - \mu I \\
 \frac{dR}{dt}&= \mu I.\nonumber
\end{align}

\begin{figure*}
 \centering
 \includegraphics[scale=0.35]{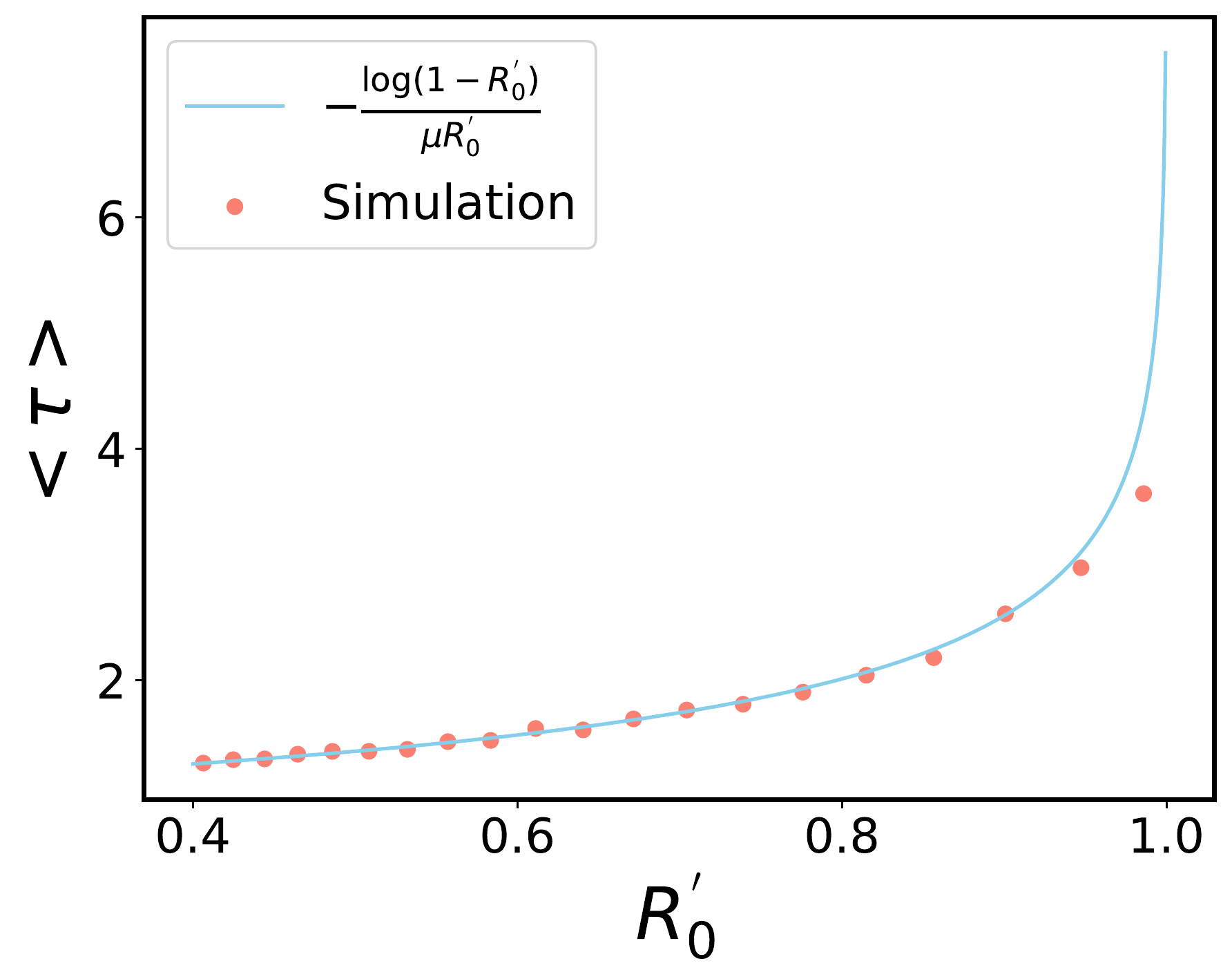}
 \caption{Average extinction time for outbreaks beyond herd immunity. Result from simulations departing from the original condition $S(0)=N(1-\alpha)-1$, $I(0)=1$ and $R(0)=N \alpha$ (continuous line) are plotted together with the analytical expression derived for sub-critical outbreaks in the SIS model [Eq. \ref{eq:T1_SIS_with_approx}]. The x-axis shows the effective reproductive number computed with $\mathcal{R}_{0}'=\frac{\beta'}{\mu}=-\mathcal{W}(-s_0\mathcal{R}_0e^{-\mathcal{R}_0})$ using $\mathcal{R}_0\in[1,2]$.}
 \label{fig:times_Super_critical_SIS_beyond_herd_immunity}
\end{figure*}

Defining the new transmission rate $\beta'=\beta (1-\alpha)$ and the new basic reproductive number $\mathcal{R}_{0}'=\frac{\beta'}{\mu}=-\mathcal{W}(-s_0\mathcal{R}_0e^{-\mathcal{R}_0})$, verifying $\mathcal{R}'_0\in(0,1)$ for $\mathcal{R}_0>1$, we mapped the super-critical SIR with initial conditions beyond herd-immunity to a sub-critical SIR in which the whole population is healthy. To proof that the regime is indeed sub-critical, remember that herd-immunity is reached when $S<\frac{\mu}{\beta}N$. This condition together with the definition of $\beta'$ ensures that $\mathcal{R}_0'<1$. In Fig. (\ref{fig:times_Super_critical_SIS_beyond_herd_immunity}) we corroborate that the formula the duration of sub-critical in the SIS also characterizes properly the super-critical outbreaks beyond herd immunity when the effective reproduction number $\mathcal{R}'_0$ is used.

\section{Duration of global outbreaks including the first wave}\label{ap:Global_outbreak_first_wave}
In order to show that the results derived in the main text are not an artifact derived from the choice of the initial conditions in Eq.~\eqref{eq:IC_recovered}, we repeat the experiments in section~\ref{sec:mobility} using initial conditions that include the first epidemic peak in the analysis.

The inclusion of mobility is especially relevant during the first wave (before herd immunity) and in the case of $\mathcal{R}_0>1$, where it is not sensible that neighboring macroscopic outbreaks can be perfectly isolated. Therefore, in this section we will study the effect of mobility in sub-populations before obtaining herd immunity. In particular, we fix
\begin{equation}\label{eq:IC_mobility}
    I_i(t=0)=R_i(t=0)=0, \quad \quad S_i(t=0)=N_i,
\end{equation}
as the initial condition for all sub-populations. An important result that we anticipate is that the mobility will accelerate the process of immunization.

\subsection{Random diffusion}

Proceeding as we did in section~\ref{subsec:diffusion}, we first focus on random diffusion between sub-populations: every agent will jump to a neighboring sub-population at constant rate $M$ ($M_{ij}=M$ if $i$ and $j$ are connected). We also consider homogeneous distributions of populations and connections. This is, the number of connections of all the populations is a constant ($k$), and the initial condition for the populations is uniform $N_i=N=10^5 \; \forall \, i \in [1,V]$. We  investigate the duration of global outbreaks with direct simulations of the process. One more time, we generate exact realizations of the process using a Gillespie algorithm, and implement a maximum time $t_{\text{max}}$ at which simulations stop.

\begin{figure}
    \centering
    \includegraphics[scale=0.35]{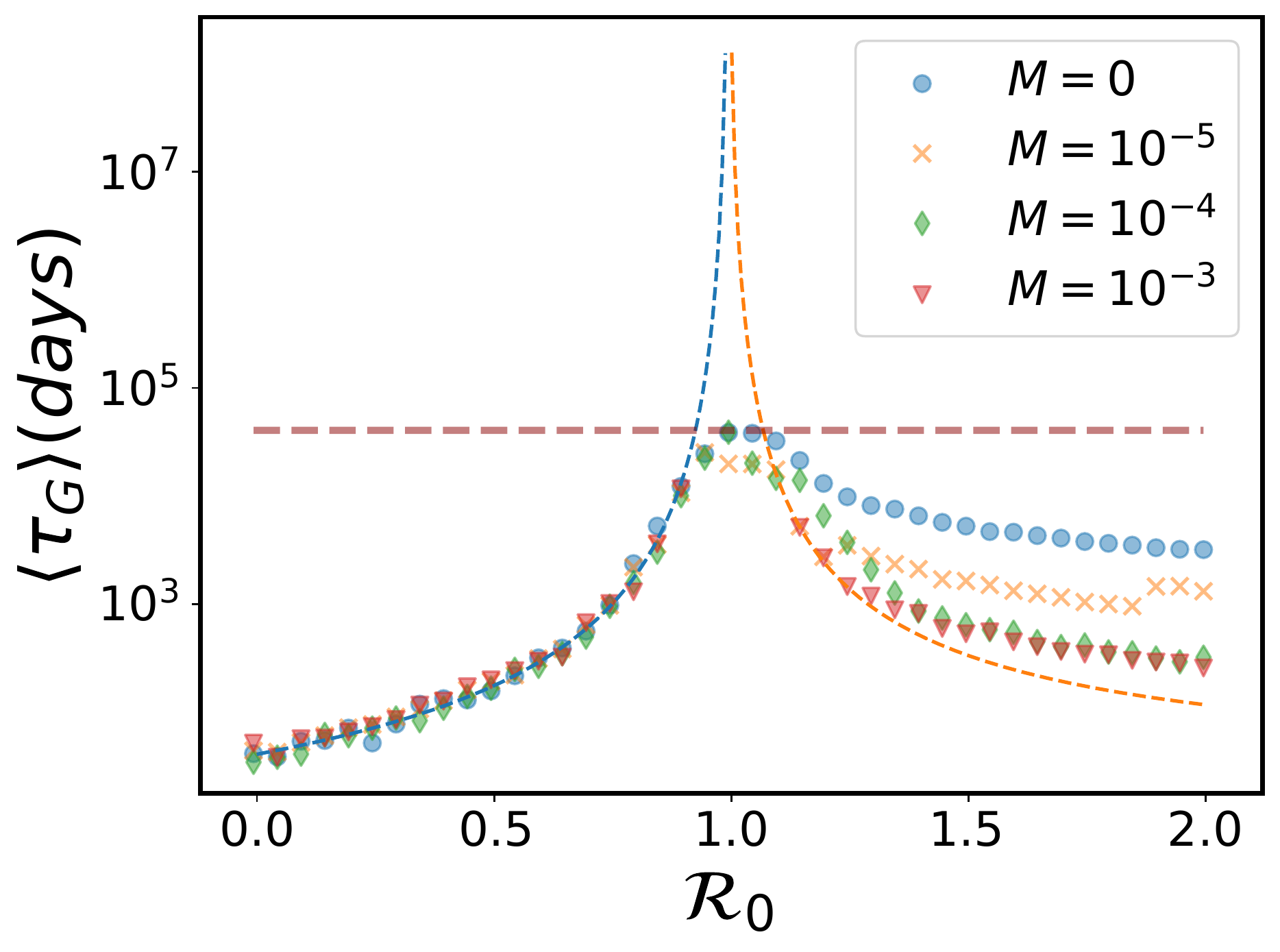}
    \caption{Effect of random diffusion on duration of global outbreaks. Dots show the duration of global outbreaks averaged over 100 simulations different values of $\mathcal{R}_0$ and $M$. The topology is a squared lattice with periodic boundary conditions ($k=4$) with $V=400$. All simulations are stopped at time $t_{\text{max}}=4\times 10^4$ (days) (horizontal dotted line). We show a transect of fixed external seeding ($h=1$ and $h/V=2.5\times 10^{-3} $). Dashed curved lines show our analytical estimations (Eqs. (\ref{eq:Analitical_average_time_V2}) for $\mathcal{R}_0 < 1$ and (\ref{eq:Analitical_average_time_super_critical}) for $\mathcal{R}_0 > 1$). As discussed in the text, the mobility doesn't have deep effect in the sub-critical phase. Nevertheless, the asynchronization of local outbreaks increase the overall duration of outbreaks for $\mathcal{R}_0>1$. This effect is reduced as the mobility rate increase.}\label{fig:Phase_diagram_mobility_diffusion}
   \end{figure}

\begin{figure}
\centering
\includegraphics[scale=0.3]{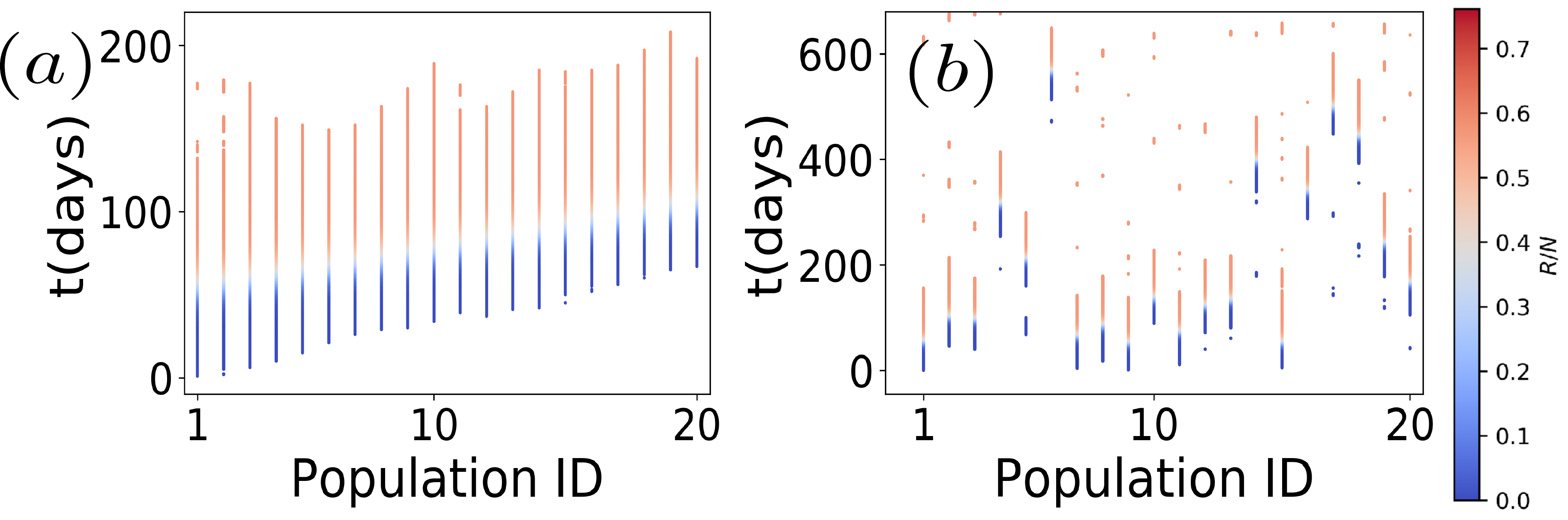}
\caption{Example highlighting the relevance of mobility on the synchronization of local macroscopic outbreaks. In (a) and (b), we plot the active sites (populations with, at least, one infected individual) for two particular realizations in a one dimensional chain with $V=20$ (extremes are not linked, $k_i=2$ for $i\in ]1,V[$ and $k_i=1$ for $i=1,V$) . In (a), we show results for $M=10^{-4}$ where a well-defined front in which populations acquired herd immunity in a synchronized fashion is clearly visible. In (d), mobility is deactivated ($M=0$), populations with a macroscopic number of recoveries (red) coexist with those in which local herd immunity is still not reached (blue). This is an instance of how asynchronous macroscopic outbreaks [in (b)] can result in overall bigger global outbreaks than those synchronized by mobility [in (a)].}
\label{fig:Synchronization}
\end{figure}

In Fig.~\ref{fig:Phase_diagram_mobility_diffusion}, we show the average duration of the first global outbreak departing from the initial condition in Eq.\eqref{eq:IC_mobility} for different values of the mobility rate $M$ and the basic reproductive number $\mathcal{R}_0$. As we expected, the effect of mobility in the sub-critical phase ($\mathcal{R}_0<1$) is negligible, and all simulations coincide with the theoretical prediction for the case of independent populations (Eq.~\eqref{eq:Analitical_average_time_V2}, shown in dashed lines). The super-critical part of Fig.~\ref{fig:Phase_diagram_mobility_diffusion} differs from the behavior described in section~\ref{sec:independent_pop}, we obtain in this case longer global outbreaks. The reason is that when departing from Eq.\eqref{eq:IC_mobility}, local outbreaks can be of macroscopic order and therefore would have longer durations (see Fig.~\ref{fig:Local_outbreak} (b)). Interestingly, even when $\mathcal{R}_0>1$, the duration of global outbreaks approaches the theoretical result derived for the case of independent populations (Eq.~\eqref{eq:Analitical_average_time_super_critical}, also shown in dashed lines) when the mobility rate increases. The reason is that mobility acts synchronizing the local macroscopic outbreaks and therefore reducing their relevance in the construction of the global outbreak. In Fig.~\ref{fig:Synchronization}, we show an example clarifying this point, it can be observed how mobility creates a well-defined front in which populations experience macroscopic outbreaks in a synchronized fashion [Fig.~\ref{fig:Synchronization} (a)]. This is in contrast with the case without mobility [Fig.~\ref{fig:Synchronization} (b)], in which macroscopic local outbreaks are  not synchronized.

\begin{figure*}
\centering
\includegraphics[scale=0.2]{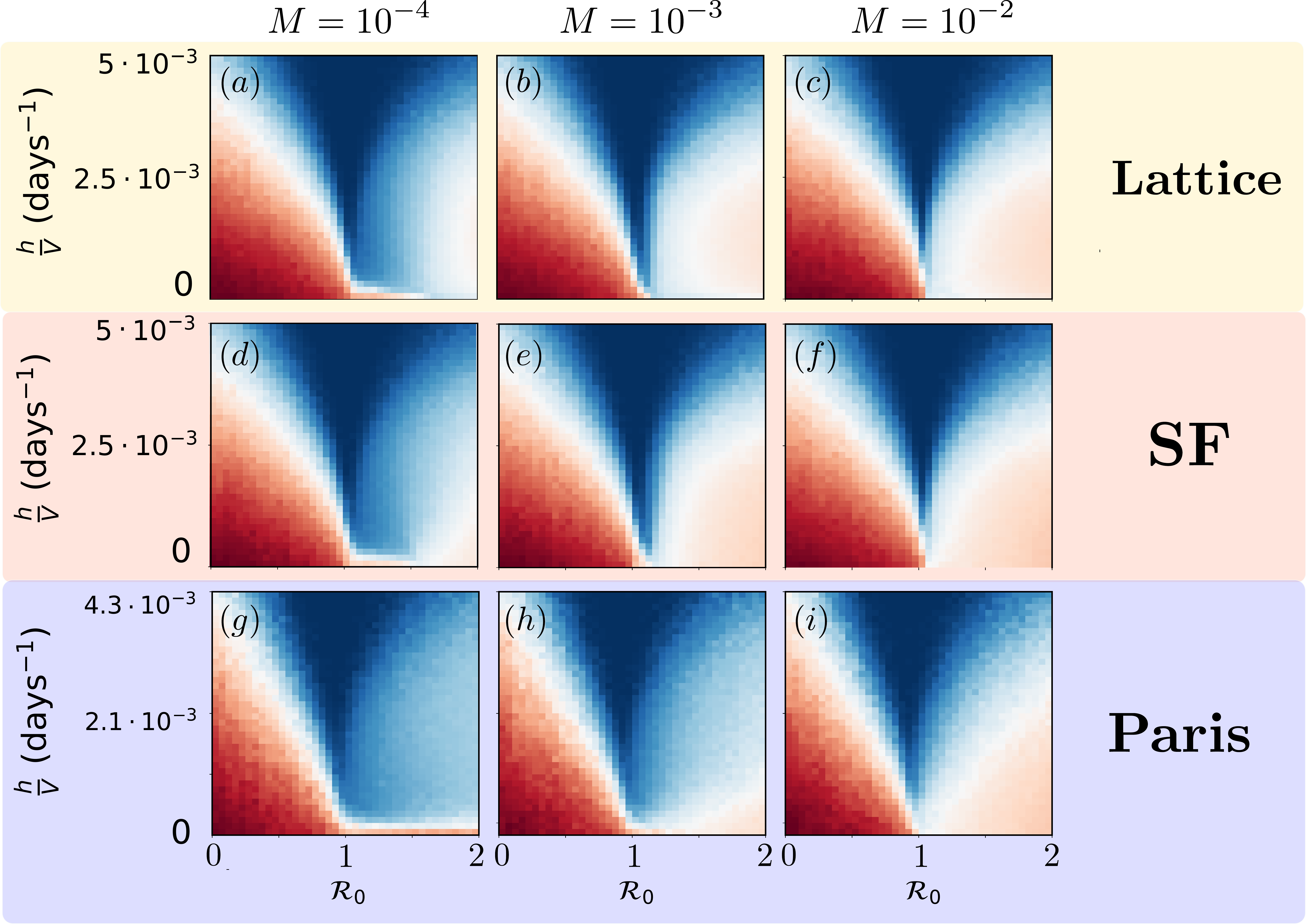} 
\caption{Average global outbreak duration for connected populations, and for different values of the portion of travelers $M$. Also,  different topologies and demographic statistics are inspected. Averages were performed over 100 realizations. Global outbreaks that do not end by the time $t_{\text{max}}=4\times 10^4$ are stopped.
In (a-c), $V=400$ populations are connected forming a regular lattice with periodic boundary conditions. Demographics are Gaussian distributed. In (d-f), the topology is scale-free network with a degree distribution $P(k) \sim k^{-2.5}$ and with $V=400$ populations proportional to the degree. In (g-i), connectivity and populations are read from commuting data of the city of Paris ($V=469$).}
\label{fig:Phase_diagram_t_first_peak}
\end{figure*}

\subsection{Recurrent mobility}

In this section we test our findings using recurrent mobility together with the initial condition in Eq.~\eqref{eq:IC_mobility}. As discussed before, this initial condition includes the possibility of having outbreaks of macroscopic magnitude in the super-critical phase. As we did in section~\ref{sec:recurrent_mobility}, we assign a sub-population of residence and work to every agent, which can be the same, and assume that they spend $1/3$ of the day in the working sub-population and the rest $2/3$ in the residence area. Mobility fluxes are one more time parameterized by the portion of agents travelling every day ($M_{i,j}$). Once the fluxes are fixed, they remain the same during all the simulation. Simulations in this setting are carried out making use of the GLEaM approximation method~\cite{balcan2010modeling}.

We demonstrate with simulations that the regimes obtained in the previous section still hold for the case of recurrent mobility and the initial condition in Eq.~\eqref{eq:IC_mobility}. Indeed, in Fig. (\ref{fig:Phase_diagram_t}), we obtain similar patterns in the phase diagram as we vary mobility $M$ for:
\begin{itemize}
 \item (a-c) A configuration in which sub-populations form a 2-D regular lattice with Gaussian distribution of the number of residents (average $10^5$ individuals and $\sigma = 3/20 \times 10^{-5}$).
 \item (d-f) Sub-populations connected by a scale-free network generated with the configurational model and with degree distribution $P(k) \sim k^{-2.5}$. The average degree is $\langle k \rangle = 6.2$. The outflow of a sub-population $i$ is equally distributed across the links connecting to it. This means that the number of individuals traveling from $i$ with population $N_i$ and degree $k_i$ to each of its neighbors $j$ is $M \, N_i/k_i$.
 \item (h-i) A realistic application in the city of Paris. The basic divisions of the city are census areas ``ensemble des communes", the resident populations and commuting networks are obtained from official statics~\cite{Demo_Paris,M_Paris}. As before, we use a control parameter $M$ to determine the fraction of resident population that commutes. The destinations are selected according to the empirical flows. For example, if $\omega_{ij}$ is the empirical number of individuals living in $i$ and working in $j$, we will consider in our simulations $M\, N_i \, \omega_{ij} / \sum_{\ell} \omega_{i\ell}$ travelers in the link $i-j$. 
\end{itemize}

\section{SEIR model}\label{ap:SEIR}
 
\begin{figure}[H]
 \centering
 \includegraphics[scale=0.5]{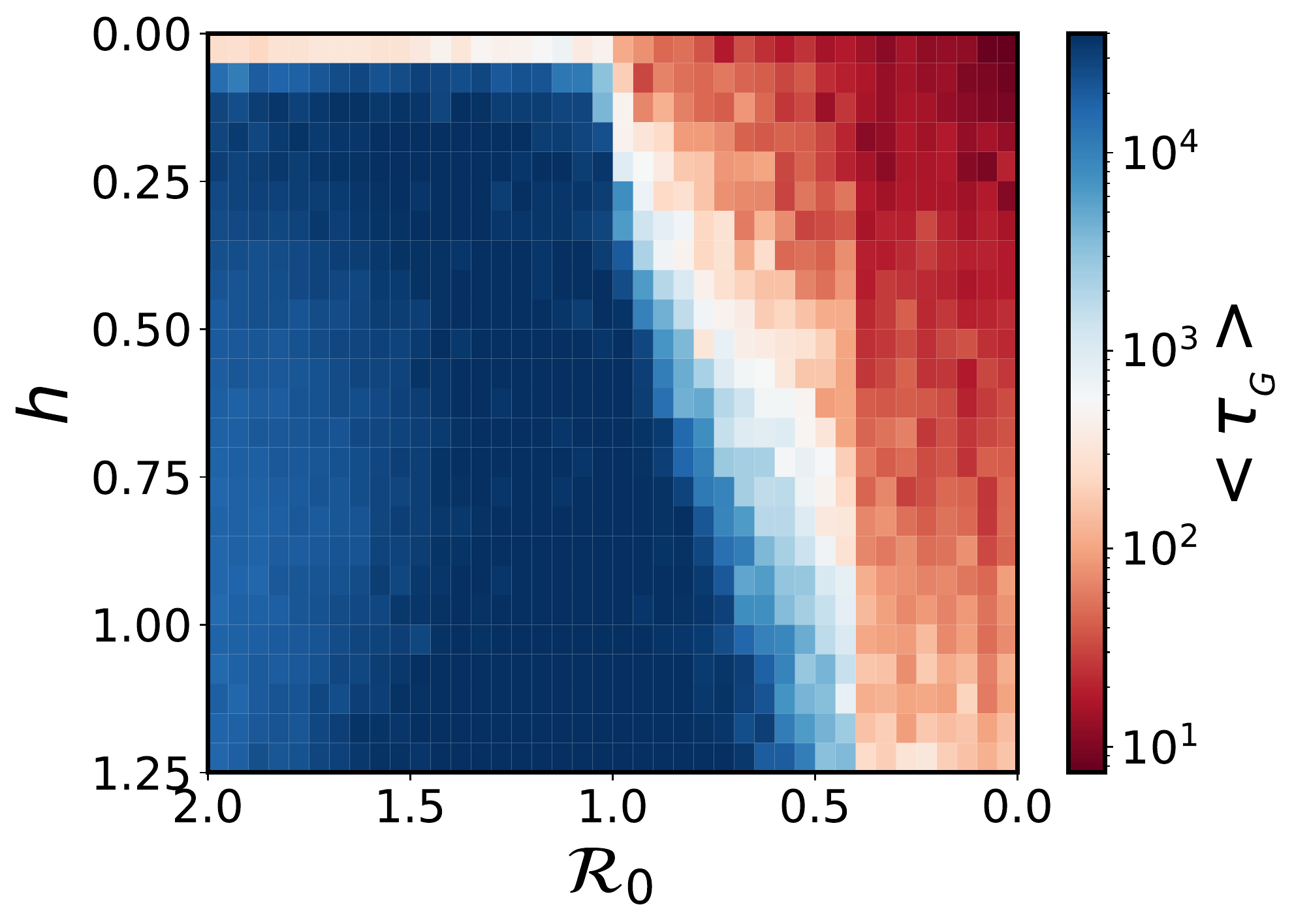}FIG
 \caption{Average duration of global outbreaks with SEIR dynamics, independent populations $(M=0)$, and for different values of $\mathcal{R}_0$ and $h$. The phase diagram is in agreement with the average duration of global outbreaks on a SIR model (Fig.~2 in the main text).} 
 \label{AP-fig:SEIR_VS_SIR_M0}
\end{figure}

\begin{figure}[H]
 \centering
 \includegraphics[scale=0.5]{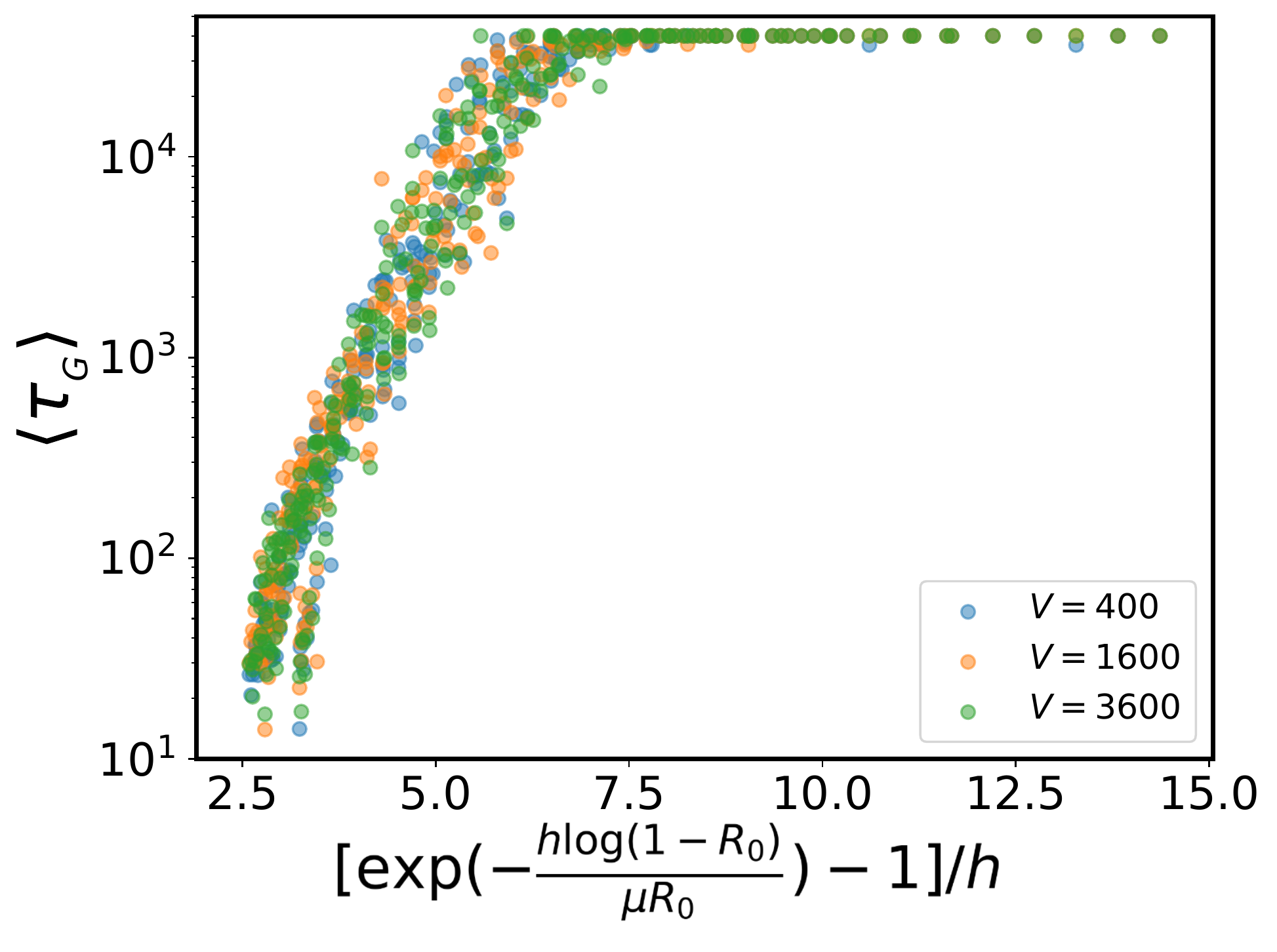}
 \caption{Scaling relation derived from Eq.~(\ref{eq:Analitical_average_time_V2}) used on SEIR dynamics with independent populations $(M=0)$}
 \label{AP-fig:Scaling_on_SEIR}
\end{figure}

 In this section, we provide evidence that the results described in the main text hold when having into account latent periods for the infection. This is, when we add a new compartment ($E$) in which agents already caught the disease but are not still infectious. Thus defining a SEIR model. In a more rigurous way, the stochastic process is defined by the following transition rates:
 \begin{align}\label{eq:macroscopic_rates_ATA_SEIR_plus_field}
    & \lim_{dt\to 0}\frac{P(E+1,I,S-1,R,t+dt|E,I,S,R,t)}{dt}=\beta \frac{I}{N} S, \nonumber \\
    & \lim_{dt\to 0}\frac{P(E-1,I+1,S,R,t+dt|E,I,S,R,t)}{dt}=\omega E, \nonumber \\
    & \lim_{dt\to 0}\frac{P(E,I-1,S,R+1,t+dt|E,I,S,R,t)}{dt}=\mu I, \nonumber
    \\
    & \lim_{dt\to 0}\frac{P(E,I+1,S-1,R,t+dt|E,I,S,R,t)}{dt}=\frac{h}{N}S , \nonumber \\
    & \lim_{dt\to 0}\frac{P(E+1,I,S-1,R,t+dt|E,I,S,R,t)}{dt}=\frac{h}{N}S , \nonumber \\
    & \lim_{dt\to 0}\frac{P(E,I+1,S,R-1,t+dt|E,I,S,R,t)}{dt}=\frac{h}{N} R.
\end{align}
One more time, we choose a characteristic scale for the average time that agents stay in the latent state ($\omega^{-1}=3.7$ days)~\cite{di2020impact}. The phase diagram is very similar to that of the SIR dynamics (see Fig.~\ref{AP-fig:SEIR_VS_SIR_M0}). But more interestingly, even the scaling deduced by Eq.~(\ref{eq:Analitical_average_time_V2}) of the main text works fairly well for the SEIR dynamics (See Figs. \ref{AP-fig:SEIR_VS_SIR_M0} and \ref{AP-fig:Scaling_on_SEIR}).

\section{Empirical study}\label{ap:data_sup}

In this section we include further information and examples of the empirical observations.

\begin{figure}[H]
 \centering
 \includegraphics[scale=0.5]{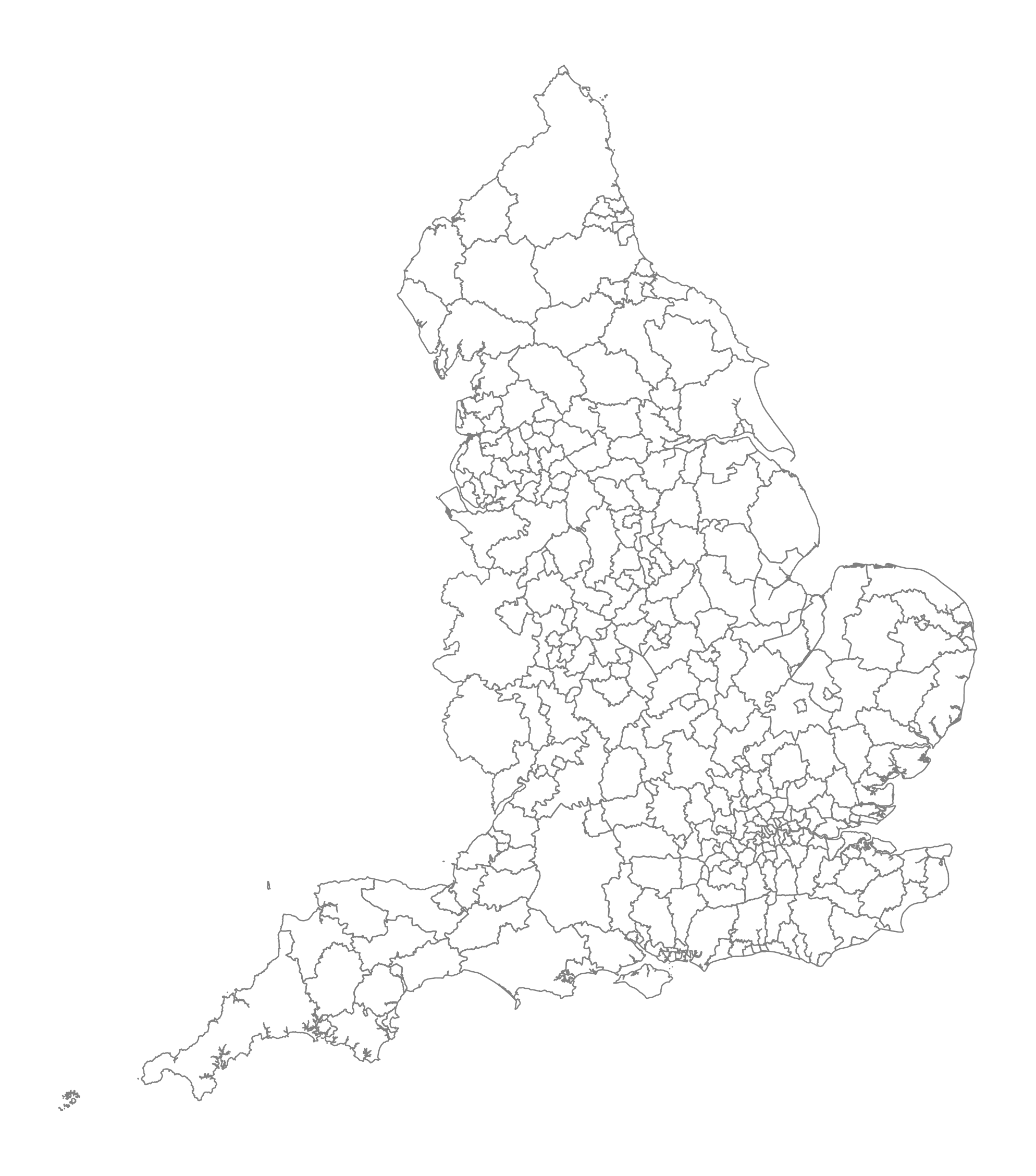}
 \caption{Geographical division of England in Lower Tier Local Authorities (LTLAs).}
 \label{AP-fig:LTLA_map}
\end{figure}

\begin{figure}[H]
 \centering
 \includegraphics[scale=0.435]{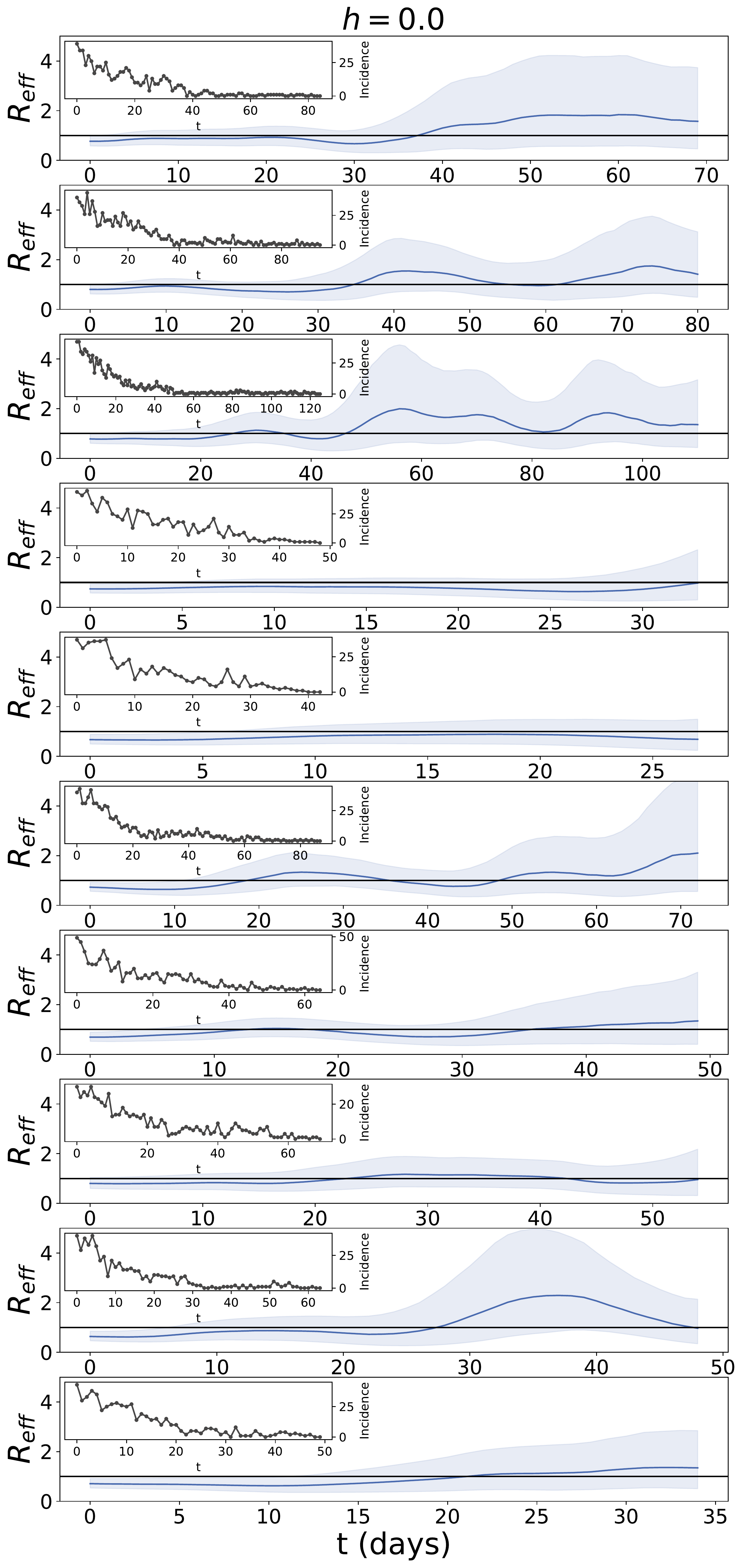}
 \caption{Each panel shows the effective reproductive number for a different realisation of the model without mobility (as described in section~\ref{sec:independent_pop} with no external field applied ($h = 0$). The inner panels are the corresponding incidence curves of each simulation.}
 \label{AP-fig:h0}
\end{figure}

\begin{figure}[H]
 \centering
 \includegraphics[scale=0.435]{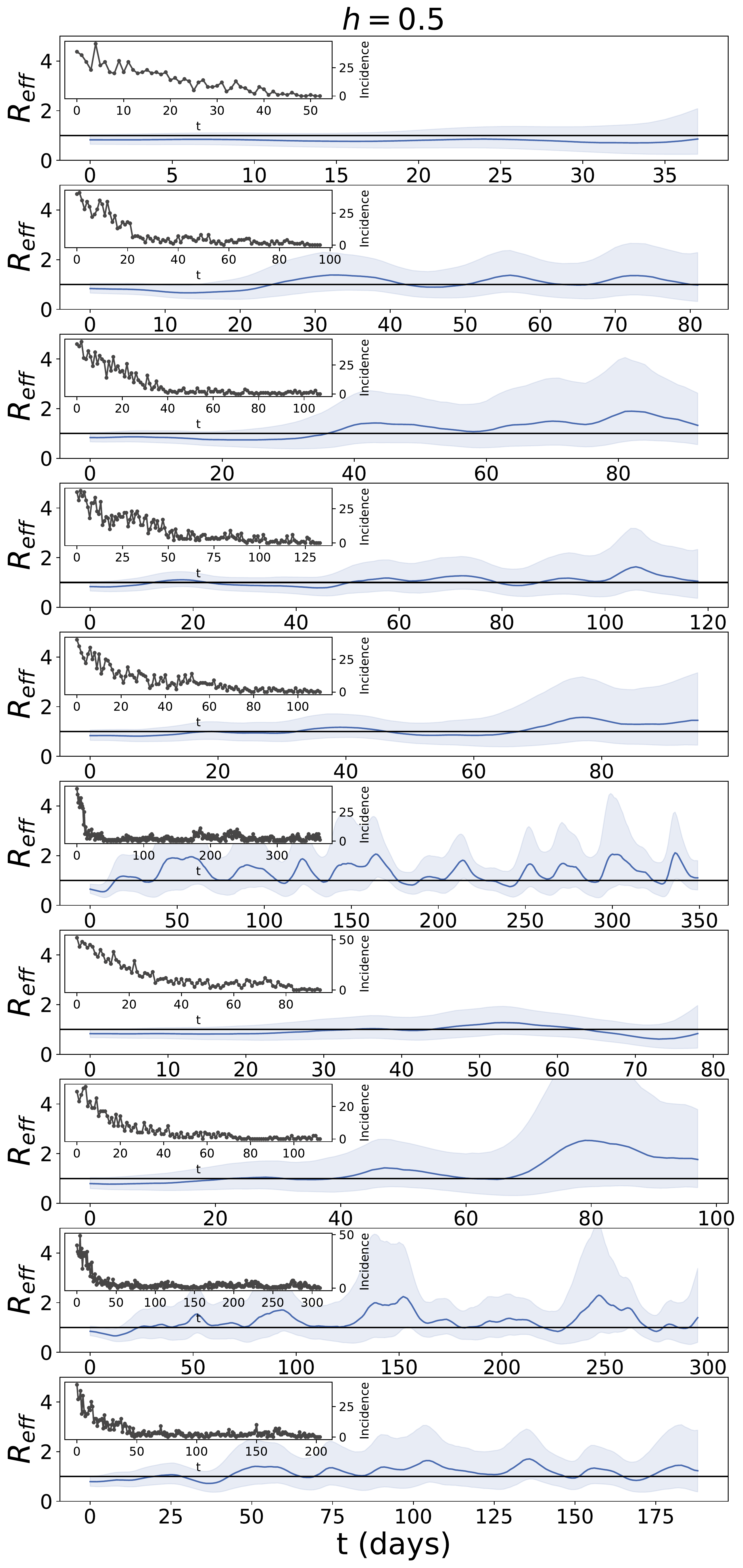}
 \caption{Each panel shows the effective reproductive number for different realisations of the model without mobility (as described in section~\ref{sec:independent_pop} with an external field $h = 0.5$. The inner panels are the corresponding incidence curves of each simulation.}
 \label{AP-fig:h05}
\end{figure}

\begin{figure}[H]
 \centering
 \includegraphics[scale=0.435]{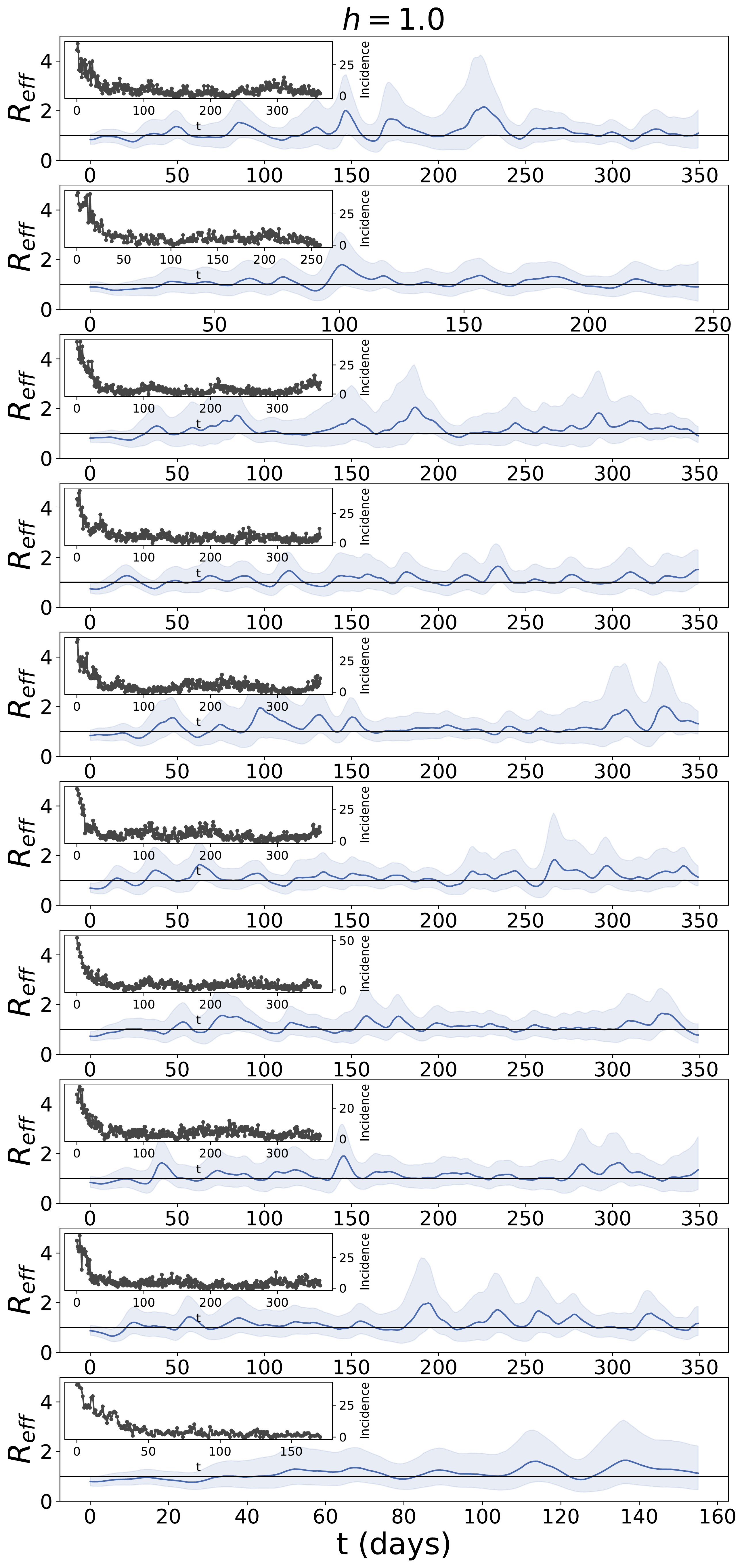}
 \caption{Each panel shows the effective reproductive number for different realisations of the model without mobility (as described in section~\ref{sec:independent_pop} with an external field $h = 1$. The inner panels are the corresponding incidence curves of each simulation.}
 \label{AP-fig:h1}
\end{figure}

The following figures show the effective reproductive number for different cases. Figure \ref{AP-fig:h0} correspond to the case where any external field is applied ($h=0$) while  in Fig. \ref{AP-fig:h05} and \ref{AP-fig:h1} an external field $h=0.5$ and $h=1.0$, respectively, is considered. In all the three cases we show simulations of the model without mobility (as described in section~\ref{sec:independent_pop}).
We are able to observe that in the $h=0$ field case the majority of the $R_{eff}$ curves have a flat and mostly subcritical profile. We can observe some cases where the incidence takes more time to vanish, resulting in more flat and fluctuating profiles  which are then reflected on oscillating values for $R_{eff}$ (Fig. \ref{AP-fig:h0}). Once we include the external field and the higher is its value, it is easier to observe fluctuating curves of the effective reproductive number. If we look in detail Fig. \ref{AP-fig:h05} we can observe different profiles of $R_{eff}$ which may be from mild to frequent oscillations between $R_{eff} = 1$ and $R_{eff} = 2$ and this again depends on the survival of the epidemic. While with the increment of the external field to $h=1.0$, Fig.\ref{AP-fig:h1}, we do not longer observe flat profiles of $R_{eff}$.

The time period we are interested in is the time between the two first waves. In the case of England we have considered the period from June 12th 2020 and August 21st 2020.

\begin{figure}[H]
 \centering
 \includegraphics[scale=0.5]{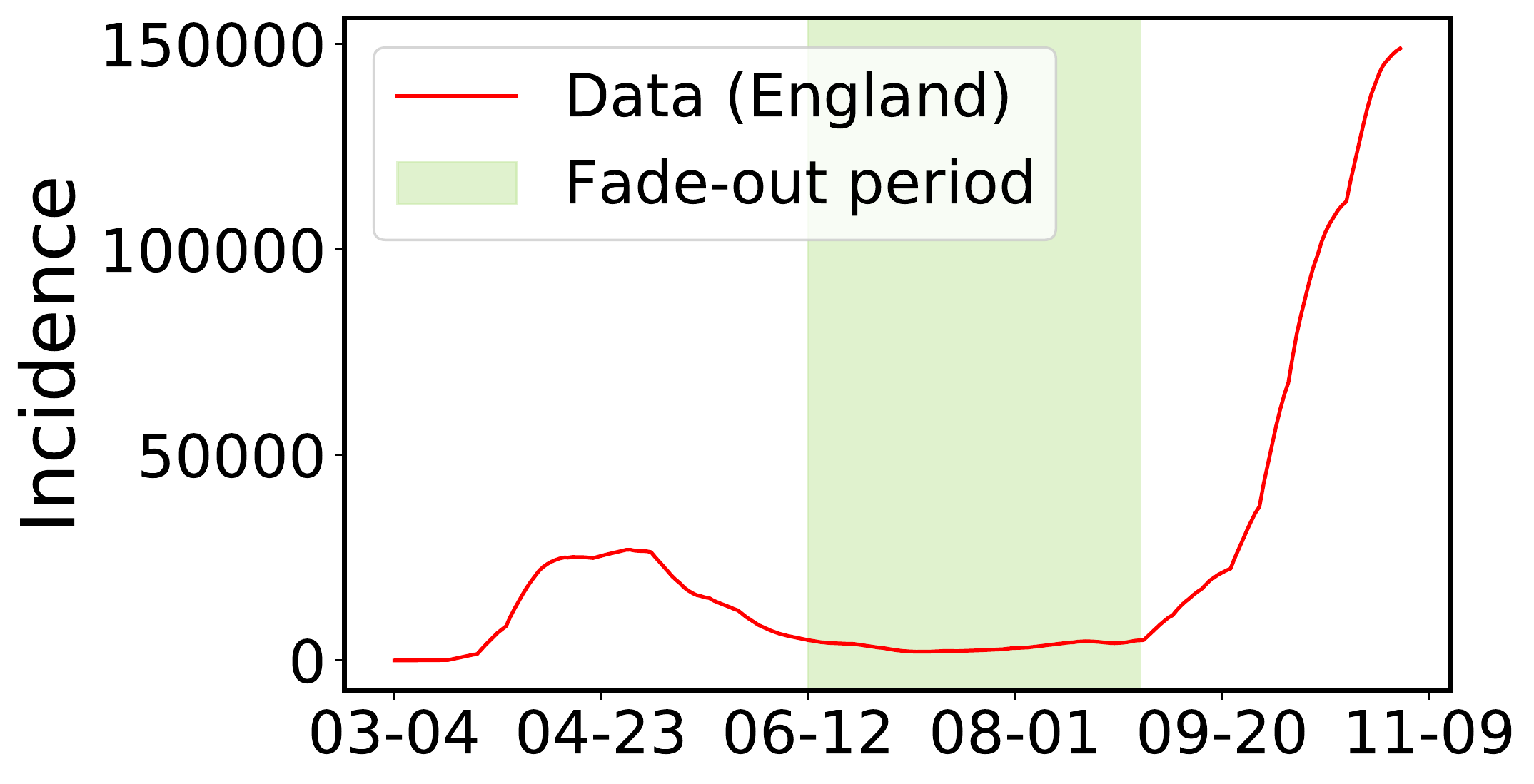}
 \caption{Incidence data in England. The green area is the period which we consider to be between the wild-type and the B.1.1.7 (Alpha) waves.}
 \label{AP-fig:inter_wave_period}
\end{figure}

\begin{figure}[H]
 \centering
 \includegraphics[scale=0.315]{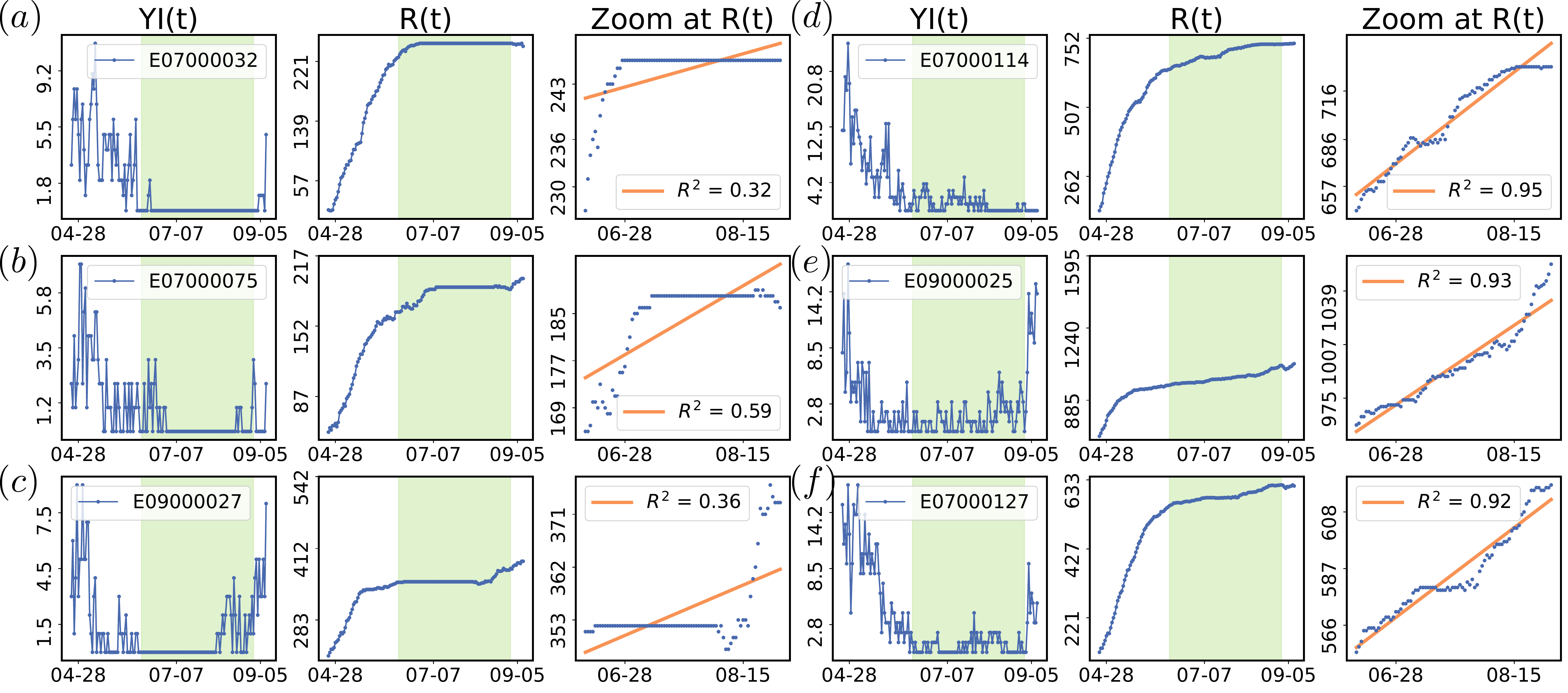}
 \caption{Left panel of each example shows the incidence of the LTLA. The central panel shows the evolution of the number of recovered individuals. Green shadows in both left and central panels are the fade-out period defined in \ref{AP-fig:inter_wave_period}. Right panel zooms over the fade-out period and we can observe a dotted blue line which is the recovered curve and an orange full line which is the linear fit of the recovered growth. (a) E07000032 is Amber Valley, local authority in Derbyshire. (b) E07000075 is Rochford, local authority in Essex. (c) E09000027 is Richmond upon Thames, local authority in London. (d) E07000114 is Thanet, local authority in Kent. (e) E09000025 is Newham, local authority in London. (f) E07000127 is West Lancashire, local authority in Lancashire.}
 \label{AP-fig:examples_linear}
\end{figure}

Figure \ref{AP-fig:examples_linear} shows an analysis of the linear growth of the number of recovered people for six different LTLAs.  

In figure \ref{fig:empirical_linear} we can see that the majority of districts are well fitted by a linear regression as we can also observe in Fig. \ref{AP-fig:examples_linear} over the examples on the right column [(d),(e), and (f)]. The districts that are well-fitted by linear functions during the studied period show low and persistent incidence. However there are also some districts, like the LTLAs on the left column [(a),(b), and (c)], which barely recorded SARS-CoV 2 incidence during summer 2020. Consequently, these cases don't show a linear growth in the number of recovered individuals.

\begin{figure}
\includegraphics[scale=0.315]{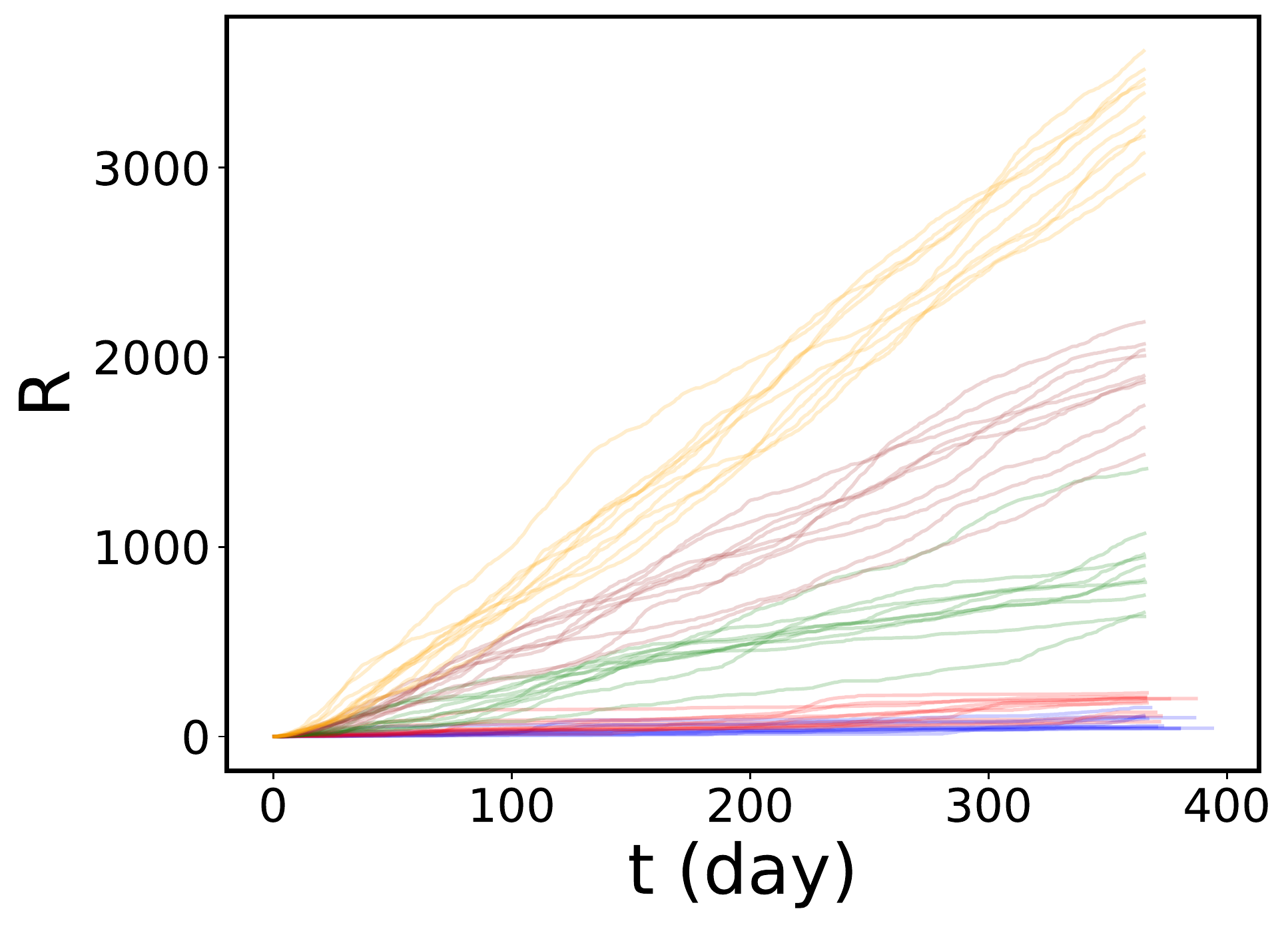}
 \caption{Examples of linear growth of the recovered individuals generated with simulations of the model. All realizations have the same initial condition $I=1$ epidemic parameters, $\mathcal{R}_0=0.8$.The external seeding varies in the interval [0.01,2].}
 \label{AP-fig:examples_linear_simulations}
\end{figure}

 

\end{document}